\documentclass[10pt]{article}

\usepackage[utf8]{inputenc}
\usepackage[T1]{fontenc}

\usepackage{amsmath, amsthm, amssymb, amsfonts, bm, mathtools}
\usepackage{mathrsfs}

\usepackage{graphicx}
\usepackage{subfigure} % consider subcaption long-term, but leaving as-is
\usepackage{booktabs}
\usepackage{multirow}
\usepackage{makecell}
\usepackage{array}
\usepackage{longtable}
\usepackage{tablefootnote}
\usepackage{caption}
\usepackage{capt-of}

\usepackage[margin=1in]{geometry}
\usepackage{parskip}
\usepackage{ragged2e}
\usepackage{hyphenat}
\usepackage{textcomp}
\usepackage{siunitx}

\usepackage[plain]{algorithm}
\usepackage{algpseudocode}

\usepackage{tikz}
\usepackage{tikz-cd}
\usetikzlibrary{automata,positioning}

\usepackage[colaction]{multicol}
\usepackage[switch]{lineno}

\usepackage[superscript]{cite}
\usepackage{mciteplus} % keep only if you really use \mcite

\usepackage{xr}

\usepackage{hyperref}

\newcolumntype{C}[1]{>{\centering\arraybackslash}p{#1}}
\usepackage{fancyhdr}
\usepackage{extramarks}

\begin{document}
\renewcommand{\thefootnote}{\fnsymbol{footnote}}

\title{TargetAgent: Mathematical AI Co-Scientist for Drug Target Identification }

\author{
Sean Cottrell$^{1}$ and  Guo-Wei Wei$^{1,2,3}$\footnote{
			Corresponding author.		Email: guowei.wei@uga.edu} \\
		\\
$^1$Department of Mathematics, \\ University of Georgia, Athens, GA  30602, USA.\\
$^2$Department of Biochemistry and Molecular Biology, \\ University of Georgia, Athens, GA  30602, USA. \\
$^3$School of Computing, \\ University of Georgia, Athens, GA  30602, USA.
}

\date{} % optional
\maketitle

% define the symbol-footnotes for the title block

	\maketitle

	\begin{abstract}
    Given the challenges of pharmaceutical development, there is increasing interest in computational approaches for identifying novel drug targets. Transcriptomics technologies enable the analysis of disease mechanisms at unprecedented resolution, while spatially resolved platforms further preserve the spatial organization of cells within tissues, allowing analysis to focus on specific regions of disease localization. To derive actionable therapeutic targets from these data, it is necessary to identify mechanisms that recur across a population rather than those observed in individual patients, which requires integrated analysis across disease conditions and donor cohorts. Here, we present the groundwork for an AI co-scientist framework for the sequential analysis of multi-batch spatial transcriptomics data that integrates interpretable mathematical models, including optimal transport, tensor factorizations, and topological data analysis, with LLM-powered agentic interpretation. Leveraging the complementary information provided by these mathematical models, the agents sequentially identify regions in which the disease is localized, molecular targets within these regions, and the direction in which each target should be modulated. Applied as a proof of concept to a public Stereo-seq study of the Alzheimer’s disease (AD)–affected prefrontal cortex, we demonstrate that the agents converge on the white matter as the primary locus of the moderate-disease signal and identify a loss of myelin programme usage despite increasing oligodendrocyte abundance. The agents subsequently propose 29 candidate therapeutic targets, led by an oligodendrocyte differentiation module on which multiple mathematical perspectives converge, and flag 20 genes whose relevance to AD has not been previously reported. These results are hypothesis-generating and motivate further validation in larger, independent cohorts.

	\end{abstract}

	\newpage

    \section{Introduction}
    The drug discovery pipeline is extremely costly in both time and money. Bringing a new drug from initial design to market approval can take upwards of a decade and over a billion dollars, and many candidates ultimately fail in clinical trials on safety or efficacy grounds \cite{dimasi2016innovation}. The identification of novel drug targets is thus an increasingly attractive direction for pharmacology \cite{pun2026target}. The task is to identify a promising therapeutic target, i.e., a specific molecule or pathway implicated in the disease, and the direction in which it should be modulated, since every downstream step of drug development inherits the mechanism fixed at this stage. This identification and targeting can either confer genuine pathological benefits to a patient, or can enable scientists to investigate the significance of different pathways in greater detail by knocking down and enhancing their activity.

    Single-cell and single-nucleus RNA sequencing have opened a new era in understanding disease pathology, with natural implications for target identification\cite{desta2024scrnaseq, mathys2019}. Spatially resolved transcriptomics goes further by preserving the tissue context of each cell or transcript, revealing patterns of cell communication and function that dissociated data cannot. Spatial technologies fall broadly into two camps. Imaging-based platforms hybridize fluorescent probes to specific transcripts and measure expression at subcellular resolution with high capture efficiency, but the targeted panel biases which genes are measured, which is not ideal for target hypothesis generation. Sequencing-based capture arrays instead combine unbiased transcript capture with positional barcodes on a structured array, at the cost of lower capture efficiency and sometimes spatial resolution \cite{stahl2016spatial}. Stereo-seq is a state-of-the-art method in this second camp: DNA nanoballs are arrayed at 500\, nm centre-to-centre spacing, so transcripts are captured at sub-micron, whole-transcriptome resolution, and a DAPI stain of the same section locates each nucleus relative to the array \cite{chen2022stereoseq}. This yields two natural views of the same tissue. Segmenting the nuclear stain (cellBin) gives a reliable proxy for single-cell expression, which can be handled with a standard scRNA-seq pipeline to determine cell types. Binning the array into 110 by 110 spot groupings (bin110) gives a 55\,$\mu$m spatial resolution identical to the popular 10X-Visium platform. At this level transcripts are dense enough to reliably delineate spatial domains and functional regions. A single analysis can therefore speak simultaneously about which cell types reside in which functional niches, which is an ideal resolution for mechanistic hypotheses.

    Omics data offer an opportunity for patient-specific biomarker analysis, but personalized drug development is not sustainable on an industrial scale. Personalized medicine can be achieved by selecting or combining drugs according to an individual's biomarkers, but target identification for drug development must instead rest on verifiable, population-level molecular mechanisms. This means finding targets that recur across donor cohorts, hence necessitates a unified analysis of many spatial datasets. That in turn raises its own challenge: diverse sources, manufacturing, and experimental conditions introduce batch effects that mask the genuine biological differences between samples \cite{tung2017batch}. Such an analysis then has to answer three questions in order: where in the tissue the disease localizes, which gene programmes define those regions and move with disease, and which genes are mechanistically important to those programmes. Each question motivates its own mathematical perspective. Mathematical approaches such as optimal transport and tensor methods have attracted increasing attention here, not only for their analytical rigor but for their interpretability \cite{bunne2024ot, cang2023commot, zeira2022paste, song2023gntd, cottrell2026storm}. Optimal transport establishes a correspondence between two measures by assigning couplings that respect a specified notion of similarity, such as similar expression activity, while minimizing a transport cost. Tensors, meanwhile, generalize matrices to higher dimensions and represent multilinear relationships among their components. Treating each spatial experiment as a lower-dimensional tensor, a multilinear factorization on the higher-dimensional tensor yields a shared spot embedding in which spatial domains are detected jointly so that a domain is comparable across donors, and a common gene programme vocabulary so that every domain is defined by its usage of the same set of pathway activities. These two approaches together yield an integration of the data that is both effective and interpretable. A transport coupling is a claim that a region of one tissue corresponds to a region of another, and unmatched mass is a claim that a region has no counterpart. A tensor factor is read as a pathway or gene programme with a spatial distribution across samples. From a pathological or pharmacological perspective, then, a programme acts through its proteins and their interactions. We therefore move to consider the programme at the level of its protein interaction structure. Topological data analysis \cite{carlsson2009topology,su2025topological}, through the persistent sheaf Laplacian \cite{wei2025sheaf}, then measures which genes are structurally essential to that programme. In practice, an interpretation of these outputs is done by hand, and it is where analyses become subjective. Large language model (LLM) agents are well suited to exactly this kind of reasoning \cite{boiko2023autonomous}, but left to themselves they are prone to conclusions that are plausible rather than verifiable \cite{ji2023hallucination}. Our premise is that the two are complementary: the mathematical models provide structured, quantitative evidence, and the agents, along with the scientific literature, provide the reasoning that turns it into novel biological hypotheses, thus enabling further scientific discovery.

    In this study, we thus propose TargetAgent: a sequential, agent-enabled mathematical pipeline built on this premise, taking multi-condition, multi-donor Stereo-seq data through from omics profiles to ranked, direction-annotated therapeutic target hypotheses. The agentic component of this pipeline is organized as a co-scientist: a panel of specialists proposes, a critic audits, and a synthesis step decides by a stated rule, with the human setting the question and judging the output. The first agent panel is asked where the disease localizes, and each of its specialists exists to consider an orthogonal perspective. A QC specialist reads sequencing depth and experimental conditions to judge batch effect and comparability across donors, so that a region is not chosen for a signal that is a property of the measurement. A composition specialist reads the cell types of each domain, so that a shift in a domain is not mistaken for a shift in its cellular make-up. A donor specialist reads the size and coherence of each domain across donors and conditions, so that the chosen region exists in every patient and can therefore support a statistical test. A transport specialist reads which domains coupled or failed to couple under the optimal transport plan, since a domain whose diseased tissue has no healthy counterpart is where the disease may have changed the tissue itself. A tensor specialist reads which tensor factor programmes are used in each domain and how that usage moves between conditions. Together these perspectives are used to propose where the disease localizes and which domains and cell types warrant further differential expression analysis. Within those domains, the therapeutic significance of a gene is then gleaned from three complementary sources: the scale of dysregulation via a DEG specialist, which can be measured from donor- and cell type-aware statistics such as NEBULA; pathway relevance, read directly off the tensor factors by a pathway specialist, which says that a gene helps define a programme that moves with disease; and the interactomic structure, measured by a topological perturbation analysis of that programme's PPI network by a topology specialist. A fourth specialist, with access to the scientific literature, determines the direction in which each gene should be modulated based on its product’s function and the direction of the biological process it regulates in disease. A second agent panel then weighs these lines of evidence against the scientific literature to identify the genes most worth pursuing or discarding.

    The novelty of the proposed agentic model pipeline is twofold. On the mathematical side, the unbalanced fused Gromov-Wasserstein couplings and the interactome are incorporated into a single graph-regularized irregular tensor factorization, so that batch smoothing, joint domain detection, and the gene programme vocabulary are products of one model rather than of three separate analyses; and the persistent sheaf Laplacian is then applied one programme at a time to ask a question that neither differential expression nor factor loadings can answer in isolation. Namely, which genes are structurally essential to a disease-associated programme. In what is increasingly referred to as 'Mathematical AI', the agents do not reason over raw expression or over the literature alone, but over the semantics of mathematical objects, each of which has a justifiable meaning: a coupling is a correspondence, unmatched mass is the absence of one, a factor is a gene programme, its usage is where the programme is active on a tissue, and a PPI positioning is a structural role. Because every claim an agent makes must name one of these quantities or a verified publication, the claims are typed and auditable; a critic can rule on each of them, and the synthesis can apply a stated rule rather than a judgment. This is what allows the pipeline to be compared against expert analysis of the same tissue, and to be held to account where the two disagree.

    \section{Methods}
    \subsection{Cell Typing of the CellBin View}
    
    The output of a Stereo-seq experiment is a bin1 matrix $X_{b_1} \in \mathbb{R}^{n_{b_1}\times n_g}$, whose $n_{b_1}$ spots lie on a 500\,nm centre-to-centre grid and whose $n_g$ columns are the genes measured. Throughout this work, we restrict to protein-coding genes by referencing against the HGNC set. Taking a cell to be roughly 15\,$\mu$m in diameter, a single cell covers on the order of 700 bin1 spots. Each experiment is accompanied by a nuclear stain image; segmenting this image approximates cell boundaries, and the spots falling inside a boundary contribute their transcripts to that cell. This gives the cellBin view of the data, $X_c \in \mathbb{R}^{n_c \times n_g}$, which can be handled with a standard single-cell RNA-seq pipeline: cells are grouped by expression similarity and the resulting clusters annotated, so that the cell-type composition of different regions of the tissue can be inferred.

    We first restrict attention to nuclei whose expression is dense enough to support a type call, retaining cells that express at least 150 distinct genes. Counts are normalized to $10^4$ per cell and log-transformed, the 2000 genes of highest dispersion are retained; expression profiles are scaled, and PCA reduces the data to $d = 50$ dimensions. For each sample $k$ this yields a representation $X_k \in \mathbb{R}^{n_{c_k} \times d}$. Harmony is then applied across the nine sections to ameliorate batch effects, and a $K$-NN graph followed by Leiden clustering groups the cells within the corrected embedding \cite{wolf2018scanpy, korsunsky2019harmony}.

    Clusters are annotated by consensus of three approaches, taking as a prior the cell types reported for the prefrontal cortex by Mathys et al. First, CellTypist transfers these labels through a pre-trained association \cite{dominguezconde2022celltypist}. Second, pseudo-bulked cluster profiles are $Z$-scored against the marker sets of that study \cite{mathys2019, gong2025stereoseq}. Third, whole-expression correlation is measured between each cluster and profiles drawn from an Allen CTX Smart-seq reference \cite{allen2019humancortex,hodge2019cortex}. A majority vote over the three assignments gives the final cell type of each cluster.

    \subsection{Spot Resolution Data Integration via Optimal Transport Enabled Irregular Tensor Factorization}
    \begin{figure}[H]
        \centering
        \includegraphics[width=\linewidth]{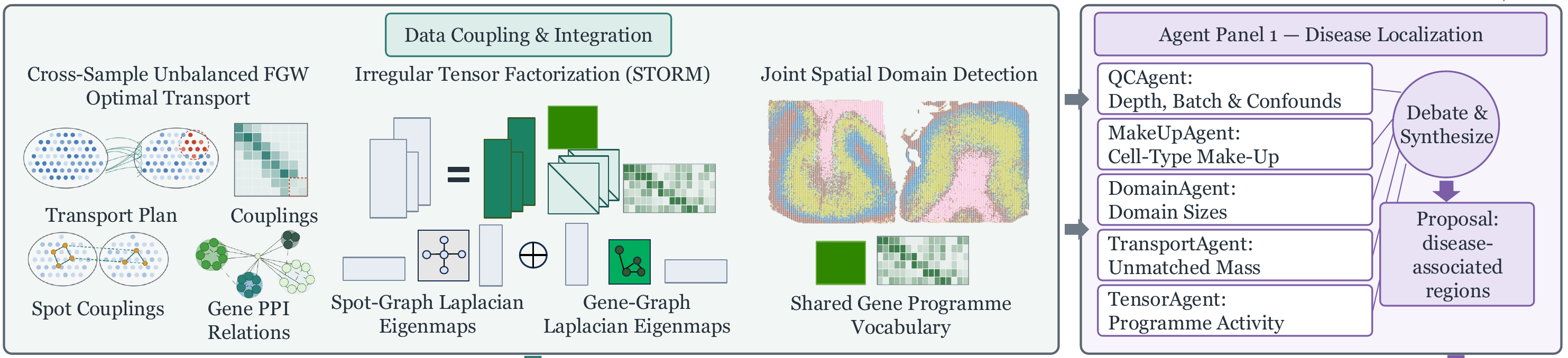}
        \caption{Cross-sample integration and the first agentic panel. Unbalanced fused Gromov-Wasserstein optimal transport yields a transport plan between the niches of each pair of joined samples, which induces spot couplings; the couplings and the within-sample spatial adjacency form a spot graph over all samples, and a PPI graph over the genes forms a gene graph. The Laplacians of the two graphs regularize the STORM irregular tensor factorization, whose shared usage embedding is clustered into joint spatial domains and whose non-negative gene factor matrix defines the shared gene programme vocabulary. The first panel then reads five perspectives on each domain: its design and quality statistics, its cell-type composition, its prevalence across donors and conditions, its matched and unmatched transport mass, and its programme usage, and proposes the disease-associated domains for investigation.}
        \label{fig:integrate}
    \end{figure}

    We now turn to the joint analysis of the multi-donor, multi-condition experiments; and to this end, we adopt the bin110 view of the data, in which bin1 profiles are aggregated into 55\,$\mu$m square tiles matching the spot size of the 10X-Visium platform. Sample $k$ of the $N$ total samples contributes an expression matrix $X_k \in \mathbb{R}^{n_{s_k} \times n_{g_k}}$ over $n_{s_k}$ spatial spots. Intersecting the protein-coding gene sets across samples gives a common set of size $n_g$, so that $X_k \in \mathbb{R}^{n_{s_k} \times n_g}$. We preprocess and rank these genes by dispersion within each sample and retain as shared highly variable genes those ranked in the top 2000 in a majority of samples.

    The multi-sample spatial transcriptomics setting is inherently multi-way: expression varies jointly across spatial location, gene identity, and sample. Concatenating the samples into a single matrix collapses this structure and obscures the geometry specific to each mode, whereas tensor decompositions provide a principled multilinear framework for such data\cite{kolda2009tensor,cottrell2026storm,song2023gntd}. We therefore represent the collection $\{X_k\}_{k=1}^{N}$ as an irregular third-order tensor $\mathcal{X}$ with frontal slices $X_k$, whose modes correspond to spot, gene, and sample. However, in spatial transcriptomics it is generally assumed that there are additional constraints on the analysis that need to be considered, such as the consideration of the tissue's spatial structure, the amelioration of batch effects between samples, and the interaction or co-expression of different genes \cite{song2023gntd,cottrell2026storm}. Graph-based regularization is a natural approach, where a network of spots within a sample can be constructed based on spatial adjacency, and a network of genes can be constructed based on interactomic information such as protein-protein interaction databases or co-expression measurements\cite{cai2011graph}.

   To ameliorate batch effects, we seek an additional correspondence between spots across different samples that can be smoothed over. Optimal Transport approaches have become a standard method to provide such a correspondence, where a transport plan is learned between measures or distributions \cite{ambrosio2002transport,peyre2019computational}. In this case, say, two sets of cells defined by their gene expression activities. However, traditional optimal transport has several limitations that make it unsuitable for this setting. First, optimal transport necessitates the preservation of mass, which means that between two distributions, all spots must find a coupling and all spots must be somehow coupled. In a disease-associated setting, there may be some regions of a tissue, such as a tumor or a plaque, that genuinely have no reasonable counterpart in healthy tissue. Unbalanced optimal transport addresses this by enabling some mass to be discarded if the cost of its transport exceeds an acceptable threshold \cite{chizat2018scaling,bryan2026trajectory}. Second, a standard optimal transport considers a cost which may be based solely on, say, gene expression similarity among spots. This view neglects the spatial arrangements of those spots, motivating the fused Gromov Wasserstein optimal transport plan where the internal geometry of each tissue is explicitly accounted for in the couplings calculations \cite{vayer2020fused,zeira2022paste}. Given such a transport plan we ultimately have a coupling of spots between tissues, which serves as a reasonable association between samples. These couplings then complete a single graph over all spots in all samples, which supplies the regularization that smooths batch and donor-induced variability in the factorization.

    \subsubsection{Unbalanced Fused Gromov Wasserstein Optimal Transport}
    Consider two samples with spots $\{x_i\}_{i=1}^{n_{s_k}}$ and
    $\{y_j\}_{j=1}^{n_{s_{\ell}}}$ carrying marginals $\mu \in \mathbb{R}_{+}^{n_{s_k}}$ and
    $\nu \in \mathbb{R}_{+}^{n_{s_{\ell}}}$. Let $C \in \mathbb{R}^{n_{s_k} \times n_{s_{\ell}}}$ be the
    ground cost, with $C_{ij}$ the dissimilarity of the expression profiles of
    $x_i$ and $y_j$, and let $D^{(1)} \in \mathbb{R}^{n_{s_k} \times n_{s_k}}$ and
    $D^{(2)} \in \mathbb{R}^{n_{s_{\ell}} \times n_{s_{\ell}}}$ hold the pairwise Euclidean spatial distances
    among spots within each tissue. A coupling
    $\Pi \in \mathbb{R}_{+}^{n_{s_{k}} \times n_{s_{\ell}}}$ is sought that minimizes
    \begin{align}
        \mathcal{J}(\Pi)
        ={}& (1-\alpha) \sum_{i,j} C_{ij}\,\Pi_{ij}
          \;+\; \alpha \sum_{i,k,j,l}
              \bigl( D^{(1)}_{ik} - D^{(2)}_{jl} \bigr)^{2}\,
              \Pi_{ij}\,\Pi_{kl} \nonumber\\
          &+ \rho_{1}\,\mathrm{KL}\!\bigl(\Pi \mathbf{1}_{n_{s_{\ell}}} \,\|\, \mu\bigr)
           + \rho_{2}\,\mathrm{KL}\!\bigl(\Pi^{\top} \mathbf{1}_{n_{s_{k}}} \,\|\, \nu\bigr)
           + \varepsilon\,\mathrm{KL}\!\bigl(\Pi \,\|\, \mu \otimes \nu\bigr).
        \label{eq:ufgw}
    \end{align}
    The first term is the Wasserstein cost, matching spots of similar
    expression; the second is the Gromov--Wasserstein cost \cite{memoli2011gromov}, which compares the
    within-tissue pairwise distances $D^{(1)}_{ik}$ and $D^{(2)}_{jl}$ over pairs of
    matched spots and thus rewards couplings that preserve each tissue's
    internal geometry, with the fusion parameter $\alpha \in [0,1]$ governing
    the trade-off. Strict mass preservation is relaxed in the
    unbalanced fashion: the hard marginal constraints
    $\Pi\mathbf{1}_{n_{s_{\ell}}} = \mu$ and $\Pi^{\top}\mathbf{1}_{n_{s_{k}}} = \nu$ are replaced by the KL penalties weighted by $\rho_{1}$ and $\rho_{2}$, so that mass whose transport is too costly is simply left uncoupled\cite{chizat2018scaling}. The final term is the entropic regularizer with weight $\varepsilon$, which renders the problem amenable to Sinkhorn-type scaling iterations for computational benefit \cite{peyre2019computational}. The mass each spot retains or loses under the optimal plan is itself informative, and the unmatched mass of each region is carried forward as evidence in the downstream agentic analysis.

    However, in spite of entropic regularization the optimal transport problem remains computationally costly when analyzing hundreds of thousands of cells across many samples. To this end, we propose several simplifying approaches to the computation. First, we coarsen the samples before coupling: spots are collected into a fixed number of spatially contiguous niches, each profiled by pooling the counts of its member spots and weighted by the number of spots it contains. The plan is then computed between niche sets rather than between individual spots. The niche plan then induces a spot-level coupling. Only spots whose niches are coupled are eligible to be joined, and within a matched niche pair each spot is connected to its nearest neighbors in the partner niche by expression similarity.

    Second, we can choose to be more targeted than a generic all-to-all correspondence in how the cross-sample couplings are introduced as regularization. Specifically, samples are joined in a chain rather than in a fully connected graph. Within a condition, samples are attached to a single anchor sample, and conditions are joined to one another by a small number of cross-anchor bridges. Cross-condition edges are the only ones along which the disease contrast is regularized, where we want tissues to be comparable across disease status without washing out genuine variability. A bridge that also spans a processing batch makes batch and diagnosis vary together. We therefore select the anchor samples joined by a bridge across conditions by those which share a chip lot, and we regularize between conditions only on this shared lot.

\subsubsection{Irregular Tensor Factorization}
Now, we turn to the tensor modeling of the gene expression datasets. As mentioned, a variable number of spots between experiments introduces irregularity to the spatial dimension of the tensor. Irregular tensor decomposition approaches such as PARAFAC2 exist primarily to deal with such a situation \cite{harshman1972parafac2, kiers1999parafac2}. To address irregularity in the number of spatial spots, each slice $X_k$ is modeled via the STORM \cite{cottrell2026storm} method through an $R$-dimensional spatial subspace, which can be naturally interpreted as microenvironment structures of the tissue, with an orthonormal basis
\[
Q_k \in \mathbb{R}^{n_{s_k} \times R},
\qquad
Q_k^\top Q_k = I_R.
\]
Collecting these slice-specific frames into a tensor as frontal slices yields
\[
\mathcal{Q} = \{Q_k\}_{k=1}^{N} \in \mathbb{R}^{n_{s_k} \times R \times N}
\]
The orthonormality constraint ensures that each $Q_k$ defines a rotation into a shared $R$-dimensional microenvironment space while preserving local geometry. Projecting expression matrices into these subspaces yields a regular spatial subspace tensor of uniform microenvironment dimension
\[
\mathcal{W} \in \mathbb{R}^{R \times n_g \times N}.
\]
The spatial transcriptomics method STORM then factorizes the subspace tensor as a Tucker Decomposition \cite{kolda2009tensor}
\[
\mathcal{W} = \mathcal{D} \times_1 H \times_2 B,
\]
where $H \in \mathbb{R}^{R \times R}$ captures shared latent gene expression covariance structure across samples, $B \in \mathbb{R}^{n_g \times R}$ contains shared gene loadings which define the universal gene programme vocabulary for all samples, and $\mathcal{D}_{::k} = D_k \in \mathbb{R}^{R \times R}$ has diagonal frontal slices and modulates sample-specific factor strengths.

In the context of gene expression data, this factorization provides a natural interpretation of the matrix and tensor components of the model. The common Gram matrix $H$ constraint ensures that all tissues share a common covariance geometry, or a common set of gene programs which define the factors. The genes which linearly comprise each gene programme are then determined by the gene loadings in $B$, on which we can impose a non-negativity constraint so that the magnitude of entries in $B$ can be read as pathway involvement of that particular gene. The diagonal matrices $D_k$ modulate the strength of activity of each gene programme in each tissue, since in some samples certain gene programs may be elevated or depressed (or there may be batch induced depth differences), such as an immune programme in a diseased tissue. Accordingly, each frontal slice of $\mathcal{W}$ satisfies a PARAFAC2 factorization
\[
W_k = H D_k B^\top.
\]
Mapping back to the original spatial coordinates then yields
\[
X_k \approx Q_k H D_k B^\top.
\]

\subsubsection{Graph-Regularized Reconstruction}

As noted, to incorporate spatial and functional priors into the tensor factorization, we introduce graph regularization jointly across spatial and gene modes. Let $n_s = \sum_{k=1}^{N} n_{s_k}$. For any tensor $\mathcal{Y} = \{Y_k\}_{k=1}^{N}$ with frontal slices $Y_k \in \mathbb{R}^{n_{s_k} \times n_g}$, define the stacking operator
\begin{equation}
\mathrm{mat}(\mathcal{Y})
=
\begin{bmatrix}
Y_1 \\
\vdots \\
Y_{N}
\end{bmatrix}
\in \mathbb{R}^{n_s \times n_g}.
\end{equation}
Functional relationships between genes are modeled using a graph with adjacency matrix
$A_g \in \mathbb{R}^{n_g \times n_g}$ (e.g., a protein–protein interaction network or a co-expression network).
Let
\[
D_g = \mathrm{Diag}(A_g \mathbf{1}),
\qquad
L_g = D_g - A_g
\]
denote the corresponding degree matrix and Laplacian of the gene-gene interactions.

 In spatial transcriptomics data analysis it is assumed that gene expression varies smoothly in spatial neighborhoods of the tissue. Within each slice, spatial adjacency is constructed using physical tissue coordinates $p_{k,i} \in \mathbb{R}^2$ (e.g., radius graph or $k$-nearest neighbors (k-NN)).
To smooth batch effects induced variation across samples, one must couple spots across the samples. Cross-sample edges are defined via the optimal transport couplings. We therefore build a block spatial Laplacian
\begin{equation}\label{eqn:spatialLap}
L_s=
\begin{bmatrix}
L_s^{(11)} & \cdots & L_s^{(1N)}\\
\vdots & \ddots & \vdots\\
L_s^{(N1)} & \cdots & L_s^{(NN)}
\end{bmatrix},
\qquad
L_s^{(k\ell)}\in\mathbb{R}^{n_{s_k}\times n_{s_\ell}}.
\tag{5}
\end{equation}
where diagonal blocks $L_s^{(kk)}$ encode within-sample spatial adjacency (threshold $\varepsilon_1$), while off-diagonal blocks
$L_s^{(k\ell)}$ encode between-sample adjacency.

For slice $k$, let $p_{k,i}\in\mathbb{R}^2$ be the physical coordinate of spot $i$.
We build a within-slice adjacency $A_s^{(kk)}\in\mathbb{R}^{n_{s_k}\times n_{s_k}}$ using a radius or $k$-NN rule, e.g.,
\[
(A_s^{(kk)})_{ij}
=
\mathbf{1}\{\|p_{k,i}-p_{k,j}\|_2\le \varepsilon_1\}.
\]

Let $z_{i,j}\in\mathbb{R}$ denote the coupling strength between spot $c_i$ in sample $k$ and spot $c_j$ in sample $\ell$. For $k\neq \ell$, we define cross-slice adjacency blocks $A_s^{(k\ell)}\in\mathbb{R}^{n_{s_k}\times n_{s_\ell}}$ by
\[
(A_s^{(k\ell)})_{ij}
=
\mathbf{1}\{z_{ij}\in \mathcal{N}_K(c^{(\ell)}_{j})\},
\]
i.e. that spot $i$ in sample $k$ is among the $K$ top couplings for spot $j$ in sample $\ell$. We then enforce an undirected graph by setting $A_s^{(\ell k)}=(A_s^{(k\ell)})^\top$.

Let $A_s$ be the block matrix with blocks $\{A_s^{(k\ell)}\}_{k,\ell}$ and define
$D_s=\mathrm{Diag}(A_s\mathbf{1})$ and $L_s=D_s-A_s$.
Then for $k\neq \ell$, $L_s^{(k\ell)}=-A_s^{(k\ell)}$ (rectangular in general),
and $L_s$ is symmetric with $L_s^{(k\ell)}=(L_s^{(\ell k)})^\top$. This defines a spatial block graph Laplacian successfully coupling all cells within each sample as well as between samples to further smooth batch effects.

The goal of graph regularization is to encourage smooth variation of reconstructed gene expression across spatial neighborhoods and functionally related genes.
Smoothness is imposed through the Cartesian product Laplacian.
For any reconstruction tensor $\mathcal{Y}$, we define the regularization  as
\begin{align}
\mathcal{R}(\mathcal{Y})
&=
\mathrm{Tr}\!\big(\mathrm{mat}(\mathcal{Y})^\top L_s\,\mathrm{mat}(\mathcal{Y})\big)
+
\mathrm{Tr}\!\big(\mathrm{mat}(\mathcal{Y}) L_g\,\mathrm{mat}(\mathcal{Y})^\top\big) \nonumber\\
&=
\mathrm{vec}(\mathrm{mat}(\mathcal{Y}))^\top
(L_g \oplus L_s)
\mathrm{vec}(\mathrm{mat}(\mathcal{Y})),
\end{align}
where
\[
L_g \oplus L_s
=
(L_g \otimes I_{n_s})
+
(I_{n_g} \otimes L_s)
\]
is the Cartesian graph product Laplacian. Recall that the reconstruction tensor is defined as
\[
\mathcal{Y} := \mathcal{Q} \Delta \mathcal{W},
\qquad
(\mathcal{Q} \Delta \mathcal{W})_{::k} := Q_k H D_k B^\top.
\]

where we use the frontal-slice operator $\Delta$ to denote a frontal slice multiplication operation between two three-way tensors.

 The graph-regularized PARAFAC2 model is formulated as
\begin{equation}
\label{eq:graph_tensor_model}
\min_{\mathcal{Q},\,\mathcal{W}}
\;
\|\mathcal{X} - \mathcal{Q} \Delta \mathcal{W}\|_F^2
+
\gamma\,\mathcal{R}(\mathcal{Y})
\quad
\text{s.t.}\quad
Q_k^\top Q_k = I_R \;\; \forall k \text{ and } \mathcal{Y} = \mathcal{Q} \Delta \mathcal{W}
\end{equation}

The resulting embeddings $\{Q_k H \}_{k=1}^{N}$ provide vertically integrated latent coordinates for each slice, which are subsequently used for downstream analyses such as joint clustering.

\subsection{Feature Selection via Statistical, Topological, and Pathway Annotation Based Methods}
\begin{figure}[H]
        \centering
        \includegraphics[width=\linewidth]{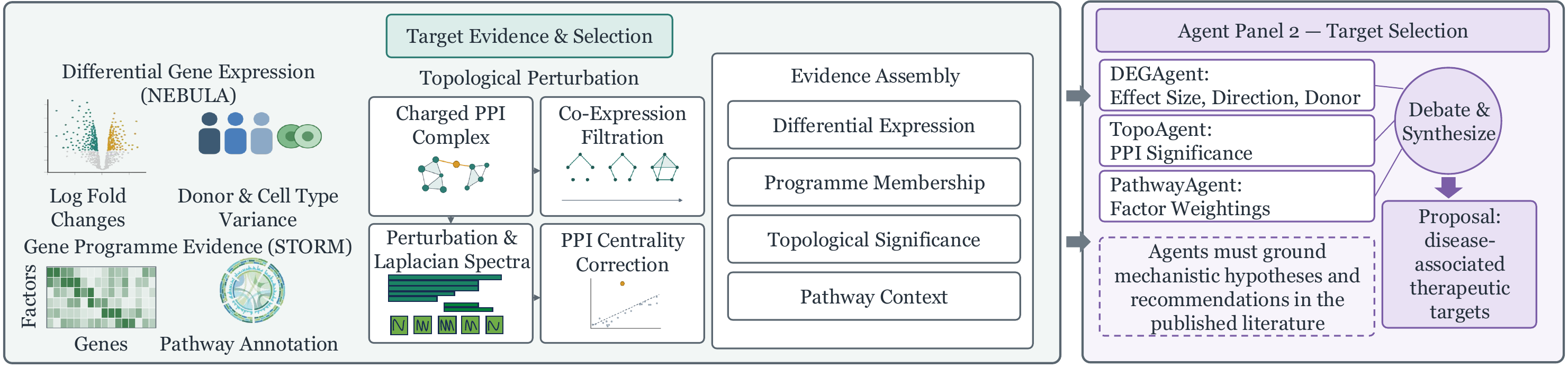}
        \caption{Target evidence and the second agentic panel. Within a shortlisted domain, three readouts are assembled for each gene: its differential expression under NEBULA, with donor and cell type entering the model; its weight in the gene programmes whose usage moves with disease, with the programmes annotated by pathway over-representation; and its topological significance, obtained for each programme whose usage moves in the domain by charging that programme's PPI clique complex with its factor weightings, filtering it by co-expression, perturbing the persistent sheaf Laplacian spectrum by removing each gene in turn, and correcting the resulting shift for the gene's weighted degree. The second panel reads the three axes in isolation, grounds mechanism and intervention direction in the published literature, and proposes therapeutic targets.}
        \label{fig:select}
    \end{figure}
We now undertake the task of identifying, given a disease-relevant cell type or spatial domain, which genes contribute to defining the disease state and which genes may be dysregulated in this region relative to healthy tissue. For pharmacological purposes, it is also important to understand how genes interact at the protein level through protein–protein interaction (PPI) analysis. The structural role of a gene can be assessed from its position within this interaction network. These goals can be addressed within three mathematical frameworks.

We start by considering differential expression across clusters and disease states. However, differential expression can be confounded by shifts in cell-type composition and variability between donors. The differential expression tool Nebula can detect statistically significant dysregulation while accounting for these other effects. Tensor factorization provides another perspective. Recall that the columns of $B$ are shared gene factors that define the spatial expression field structure across all samples. The rows of $B$ represent the contribution of each gene to each factor. This provides a natural interpretation in terms of variation in gene programs or factor activity, which can be further interpreted through pathway annotation of each factor.

For the PPI analysis, we utilize tools from Topological Data Analysis to understand the higher order structure of the PPI networks. Hub gene analysis in interactomic analysis is traditionally limited in handling only the low-dimensional, or pairwise, relations between genes, which may result in an incomplete understanding of the complex, high-dimensional, non-linear nature of gene-gene interactions. Additionally, a hub that is central to generic PPI interactions may not necessarily be a central player in the disease-associated pathway. A tensor factor which defines the disease-associated domain carries a score per gene, which can be used to label the PPI network for additional non-geometric information regarding significance. This can be viewed rigorously via cellular sheaves on the simplicial complex. The persistent Sheaf Laplacian then is an operator which can support a topological analysis of this network and its disease-associated status.

\subsubsection{Differential Gene Expression Analysis}
    Differential expression was tested at the level of individual nuclei within each selected domain, using NEBULA \cite{he2021nebula}, a negative-binomial mixed model in which the donor section enters as a random intercept and each nucleus's library size as an offset. Per-section sequencing depth was included as a fixed covariate, since depth differs systematically between the samples. Whole-domain contrasts additionally carry cell type in the design, to ensure that a group effect is not simply a change in cellular composition, stratified contrasts test one cell type at a time and so attribute a hit to a population. P-values were adjusted by the Benjamini-Hochberg procedure \cite{benjamini1995fdr} within each contrast, and a gene was called only if it reached an adjusted p below 0.05 with an absolute log2 fold change above one.

\subsubsection{Factor Matrix Pathway Annotation}
Each sample's spot-gene matrix $X_k$ is approximated in STORM via $Q_k H D_k B^T$. The gene factor matrix $B \in \mathbb{R}^{n_g \times R}$ is shared by every sample and constrained to have non-negative entries; its entry $B_{gr}$ is the contribution of gene $g$ to programme $r$, so that each column defines a programme as a set of co-varying genes on each tissue. $Q_k H$ is the usage matrix, with one row per spot giving that spot's loading on each programme, and $D_k$ is a diagonal matrix of non-negative per-sample amplitudes, one value per programme per sample. Their product $C_k = Q_k H D_k$ is the activity matrix, whose $r$-th column is the expression field of programme $r$ on the tissue of sample $k$. We have
\begin{equation}
(Q_k H)^T (Q_k H) = H^T H, \qquad C_k^T C_k = D_k H^T H D_k,
\end{equation}
which is the PARAFAC2 constraint. The programmes therefore co-occur to the same extent in every tissue, up to the rescaling of each programme by its amplitude, and the total usage of each programme is fixed within every sample. 

This structure determines which genes are highly loaded in which factor. With the remaining factors held fixed the update for $B$ in the graph-regularized model is
\begin{equation}
B^T = \Bigl[\bigl(1+\tfrac{\mu}{2}\bigr)\, G\Bigr]^{-1} \sum_{k=1}^{N} C_k^T \Bigl( X_k + \tfrac{\mu}{2}\,(Y_k + U_k) \Bigr), \qquad G = \sum_{k=1}^{N} C_k^T C_k = \sum_{k=1}^{N} D_k H^T H D_k,
\end{equation}
where $Y_k$ is the $k$-th slice of the auxiliary tensor $\mathcal{Y}$ on which the graph regularization acts, $U_k$ is the scaled Lagrange multiplier enforcing $\mathcal{Y} = \mathcal{Q}\Delta\mathcal{W}$, and $\mu > 0$ is the penalty parameter. The quantity being fitted is thus the raw expression together with the graph-smoothed expression and its multiplier. The term $C_k^T(\cdot)$ measures how strongly the expression map of each gene overlaps the expression field of each programme, pooled over all samples, and $G^{-1}$ corrects these overlaps for the size of each field and for the overlap among fields, which is governed by the shared Gram matrix $H^T H$. The non-negativity constraint is enforced within the update by exact column-wise block coordinate descent, in which each entry is the closed-form minimizer of a scalar quadratic clipped at zero, so that a negative association receives a loading of exactly zero.

Let $c_r$ denote the $r$-th column of the stacked activity matrix $[C_1; \dots; C_N]$, that is, the expression field of programme $r$ over the spots of all samples, and let $x_g$, $y_g$, and $u_g$ denote the $g$-th columns of the correspondingly stacked $X_k$, $Y_k$, and $U_k$. The loading of gene $g$ on programme $r$ is then
\begin{equation}
B_{gr} = \frac{1}{1+\tfrac{\mu}{2}} \; \frac{\bigl\langle M_r c_r,\; x_g + \tfrac{\mu}{2}\,(y_g + u_g) \bigr\rangle}{\lVert M_r c_r \rVert^2},
\end{equation}
where $M_r$ is the orthogonal projector onto the complement of the span of $\{c_s : s \neq r\}$. The loading is therefore a measure of the orthogonal importance of the gene to the programme: it is the regression of the gene's expression onto the part of that programme's expression field which no combination of the other programmes can reproduce. Since the expression enters as a sum, a gene is highly loaded when it contributes to the field through its raw expression, through its graph-smoothed expression, or through both. In this sense the highest-loading genes of a programme are those that differentially define its region of the expression map. 

Programmes were then annotated from these loadings in two ways to infer biological function. Cell-type identity was assessed by dividing the mean loading of each prior marker set by the mean loading of all genes, which assigns a programme to a lineage where one set stands out. Functional annotation was obtained by testing the fifty highest-loading genes of each programme for over-representation with a hypergeometric test, taking the variable genes identified in the corresponding dataset as the background set, against Gene Ontology Biological Process, Reactome, KEGG, PanglaoDB cell-type markers, and DisGeNET disease-gene sets \cite{kuleshov2016enrichr,franzen2019panglaodb,pinero2020disgenet}.

Finally, conditions were compared on centred usage. Each section is thus its own reference, so that any section-wide influence, whether sequencing depth or cell-type composition, cancels, and the quantity tested is whether a programme is used more in a given domain than in the rest of the same tissue. A programme that defines a domain but whose centred usage there falls in diseased relative to control samples indicates a disease-associated shift in that domain.

\subsubsection{Simplicial Homology}
    We note, however, that a gene which defines the orthogonal information of a programme is not necessarily the gene which is maximally important for the functioning of a specific pathway. To this end, we must then consider how genes interact at the level of their protein products. This motivates a topological analysis of the PPI network comprised of the genes from each disease associated tensor factor. To introduce the topological analysis of the PPI network, we first briefly review the notion of a simplicial complex and combinatorial Laplacian. We first define a simplex, simplicial complex, $q$-chain, and boundary. A 0-simplex is a vertex, a 1-simplex is an edge, a 2-simplex is a triangle, and so on. Generally, we consider a $q$-simplex, $\sigma_q$. A simplicial complex is then a means of approximating a topological space by gluing together the faces of simplices. More formally, a simplicial complex $K$ is a collection of simplices such that:
\begin{enumerate}
    \item If $\sigma_q \in K$ and $\sigma_p$ is a face of $\sigma_q$ then $\sigma_p \in K$.
    \item The nonempty intersection of any two simplices is a face of both simplices.
\end{enumerate}
A $q$-chain is then defined as a sum of $q$-simplices taken from the simplicial complex $K$ with coefficients in some field, such as $\mathbb{Z}_2$. The set of $q$-chains naturally has a basis in the set of $q$-simplices in $K$, and this set forms a finitely generated free Abelian group $C_q(K)$. We then define the boundary operator as a homomorphism relating the Chain groups, $\partial_q : C_q(K) \rightarrow C_{q-1}(K)$. The boundary operator is defined as:
\begin{equation}
    \partial_q \sigma_q = \sum_{i=0}^q(-1)^i\sigma^i_{q-1}.
\end{equation}
where $\sigma^i_{q-1}$ is a $q-1$ simplex obtained via the removal of the $i$'th vertex from $\sigma_q$. The sequence of chain groups connected by this homomorphism is then a Chain Complex:
\begin{center}
    $... \xrightarrow{\partial_{q+1}} C_{q}(K) \xrightarrow{\partial_{q}} C_{q-1}(K) \xrightarrow{\partial_{q-1}} ...$.
\end{center}
It is well known that the boundary operator and the Chain Complex associated with a simplicial complex gives the number of $q$-dimensional holes in that topological space. Specifically, the $q$th Homology Group is defined as $H_q = \text{ker} \partial_q / \text{Im} \partial_{q+1}$. Equivalently, it is the collection of chains with no boundary who are not themselves the boundary of another higher dimensional chain. The dimension of this group is also known as the $q$th Betti Number, $\beta_q$. The matrix representation of the $q$th boundary operator with respect to the standard basis of simplices in $C_q(K)$ and $C_{q-1}(K)$ is denoted as $\mathcal{B}_q$. We can then equip all chain groups with the canonical inner product and define the adjoint operator

\begin{equation}
    \partial_q^*: C^{q-1}(K) \rightarrow C^q(K),
\end{equation}
which in matrix form is given as the transpose of $\mathcal{B}_q$, denoted $\mathcal{B}_q^T$, with respect to the same basis. We can now define the $q$-combinatorial Laplacian matrix as:
\begin{equation}
    \mathcal{L}_q := \mathcal{B}_{q+1}\mathcal{B}_{q+1}^T + \mathcal{B}_q^T\mathcal{B}_q.
\end{equation}
The harmonic spectrum of the $q$-combinatorial Laplacian matrix reveals the dimension of the $q$th Homology group, or the number of $q$-dimensional holes in our simplicial complex. The non-harmonic spectrum then reveals further geometric shape information \cite{wang2020persistent}. Intuitively, $\beta_0$ reveals the number of connected components in $K$, $\beta_1$ reveals the number of loops in $K$, and $\beta_2$ reveals the number of 2D voids in $K$.
    \subsubsection{Cellular Sheaves on a Labeled Simplicial Complex}
    Cellular sheaf theory has previously been defined over networks and simplicial complexes \cite{wei2025sheaf,hansen2019sheaves}. A cellular sheaf on a simplicial complex $X$ consists of the following:
    \begin{enumerate}
        \item A simplicial complex $X$, where the face relation that $\sigma_p$ is a face of $\sigma_q$ is denoted by $\sigma_p \leq \sigma_q$
        \item An assignment to each simplex $\sigma$ of $X$ a finite dimensional vector space $\mathcal{F}(\sigma)$ and to each face relation $\sigma_p \leq \sigma_q$ a linear morphism of vector spaces denoted $\mathcal{F}_{\sigma_p \leq \sigma_q}:\mathcal{F}(\sigma_p) \rightarrow \mathcal{F}(\sigma_q)$ satisfying $\sigma_r \leq \sigma_p \leq \sigma_q \implies \mathcal{F}_{\sigma_r \leq \sigma_q} = \mathcal{F}_{\sigma_r \leq \sigma_p} \circ \mathcal{F}_{\sigma_p \leq \sigma_q}$ and $\mathcal{F}_{\sigma_p \leq \sigma_p}$ is the identity map.
    \end{enumerate}
    The vector space $\mathcal{F}(\sigma)$ is the stalk over the simplex $\sigma$ and the linear morphism $\mathcal{F}_{\sigma_p \leq \sigma_q}$ is the restriction map of the face relation between the two simplices. A global section $s$ of $\mathcal{F}$ is an assignment to each simplex $\sigma$ an element $s_\sigma \in \mathcal{F}(\sigma)$ such that  $\mathcal{F}_{\sigma_p \leq \sigma_q}(s_{\sigma_p}) = s_{\sigma_q}$ for the restriction map associated to any face relation $\sigma_p \leq \sigma_q$.

    Suppose then that we have a graph, or a one dimensional simplicial complex $X$ where each vertex $v_i$ is associated with a quantity $q_i \in \mathbb{R}$. Denote the edge connecting $v_i$ and $v_j$ as $e_{ij}$. We can define a sheaf $\mathcal{F}$ on $K$ such that each stalk is one-dimensional $\mathbb{R}$, and for the face relation $v_i \leq e_{ij}$, the morphism  $\mathcal{F}_{v_i \leq e_{ij}}$ is the multiplication by $q_j / r_{ij}$ where $r_{ij}$ is the length of $e_{ij}$. The assignment $q_i \rightarrow v_i$ and $q_iq_j/r_{ij} \rightarrow e_{ij}$ is then a global section, since $\mathcal{F}_{v_i \leq e_{ij}}(q_i) = \mathcal{F}_{v_j \leq e_{ij}}(q_j) = q_iq_j/r_{ij}.$

    This can then be generalized to higher order simplicial complexes. Suppose we also have a nowhere zero function $F: X \rightarrow \mathbb{R}$. We can define a sheaf where each stalk is $\mathbb{R}$ and for the face relation $[v_0, v_1, ..., v_n ] \leq [v_0,v_1,..,v_n,v_{n+1}, ..., v_m ]$, the linear morphism $\mathcal{F}([v_0, v_1, ..., v_n ] \leq [v_0,v_1,..,v_n,v_{n+1}, ..., v_m ])$ is the scalar multiplication by:
    \begin{center}
        $\frac{F([v_0, v_1, ..., v_n])q_{n+1}\cdot \cdot \cdot q_m}{F([v_0, v_1, ..., v_n,v_{n+1}, ..., v_m])}$
    \end{center}
     this, similarly, can provide a global section.

    \subsubsection{Persistent Sheaf Laplacian}
    The Sheaf Laplacian is a combinatorial Laplacian constructed from the sheaf cochain complexes. The sheaf cochain complex of $\mathcal{F}$ can be defined as follows. Let the q-th cochain group $C_q(X,\mathcal{F})$ be the direct sum of $\mathcal{F}(\sigma)$ over all q-simplices in $K$. To define the coboundary map $d$, we need to account for orientation via a signed incidence relation, which is an assignment to every face relation $\sigma_p\leq \sigma_q$ an integer $[\sigma_p : \sigma_q]$ satisfying the following:
    \begin{center}
        \begin{enumerate}
            \item If dim $\sigma_q$ - dim $\sigma_p$ $>$ 1, then $[\sigma_p : \sigma_q]$ = 0
            \item If $\sigma_r \leq \sigma_q$ and dim $\sigma_q$ - dim $\sigma_r$ = 2, then $\sum_{\sigma_p} [\sigma_r : \sigma_p][\sigma_p : \sigma_q] = 0$
        \end{enumerate}
    \end{center}
    If a signed incidence relation is given, we then define the coboundary map $d^q : C_q(X;\mathcal{F}) \rightarrow C_{q+1}(X;\mathcal{F})$ by
    \begin{center}
        $\partial^q |_{\mathcal{F}(\sigma_p)} = \sum_{\sigma_p \leq \sigma_q}[\sigma_p : \sigma_q] \mathcal{F}_{\sigma_p \leq \sigma_q}$
    \end{center}
    Which induces a sheaf cochain complex
    \begin{center}
        $0 \rightarrow C_0(X,\mathcal{F})\rightarrow C_1(X,\mathcal{F})\rightarrow C_2(X,\mathcal{F}) \rightarrow ...$
    \end{center}
    The (q+1)-th sheaf cohomology group $H_q(X,\mathcal{F})$ is defined by ker $\partial^{q+1}$/im $\partial^{q}$. Persistent sheaf cohomology is known to many researchers. Suppose that $f : X \rightarrow Y$ is a simplicial map and that $\mathcal{F}$ is a cellular sheaf on $Y$. The pullback sheaf $f \ast \mathcal{F}$ on $X$ is given by
    \begin{center}
        $(f \ast \mathcal{F})(\sigma) = \mathcal{F}(f(\sigma))$
    \end{center}
    and for the face relation $\sigma_p \leq \sigma_q$ of $X$:
    \begin{center}
        $(f \ast \mathcal{F})_{\sigma_p \leq \sigma_q} = \mathcal{F}_{f(\sigma_p) \leq f(\sigma_q)}$
    \end{center}
    Given two oriented simplicial complexes $K, Y$, if $K \subset Y$ and the orientation of $K$ is identical to $Y$, let sheaf $\mathcal{F}$ on $K$ be the pullback of the sheaf $\mathcal{G}$ on $Y$,then we have the following commutative diagram:
    \begin{center}
    \begin{tikzcd}
    \cdots \arrow[r, "\partial"]
    & C^{q-1}(K; \mathcal{F}) \arrow[r, "\partial"]
    & C^q(K; \mathcal{F}) \arrow[r, "\partial"]
    & C^{q+1}(K; \mathcal{F}) \arrow[r, "\partial"]
    & \cdots \\
    \cdots \arrow[r, "\partial"]
    & C^{q-1}(Y; \mathcal{G}) \arrow[r, "\partial"] \arrow[u, "\pi"]
    & C^q(Y; \mathcal{G}) \arrow[r, "\partial"] \arrow[u, "\pi"]
    & C^{q+1}(Y; \mathcal{G}) \arrow[r, "\partial"] \arrow[u, "\pi"]
    & \cdots
    \end{tikzcd}
    \end{center}
    where $\pi : C^q(Y, \mathcal{G})\rightarrow C^q(K, \mathcal{F})$ is a projection map. Since $\pi$ is a cochain map, it induces a map $\pi_*$ between sheaf cohomology groups of $\mathcal{F}$ and $\mathcal{G}$, and the q-th persistent sheaf homology group is defined by:
    \begin{center}
    $\pi_*(H^q(Y,\mathcal{G})$
    \end{center}
    whose dimension is the q-th persistent sheaf Betti number. To extend the notion of the persistent Laplacian to cellular sheaves, we must dualize the above diagram. Note that the adjoint of $\pi$ is the inclusion map $\iota : C^q(K,\mathcal{F}) \rightarrow C^q(Y,\mathcal{G}).$
    \begin{center}
\begin{tikzcd}
    C^{q-1}(K; \mathcal{F}) \arrow[rr, "d^{q-1}_X", shift left] %\arrow[dd, hook, dashed]
     &
      & C^{q}(K; \mathcal{F}) \arrow[ll, "(d^{q-1}_K)^{\ast}", shift left] \arrow[rd, "d^{q}_{K,Y}", shift left] \arrow[dd, hook, dashed]
       &
        & \\
     &
      &
       & \Theta^{q+1}_{K,Y} \arrow[lu, "(d^{q}_{K,Y})^{\ast}", shift left] \arrow[rd, hook, dashed]
        & \\
    %C^{q-1}(Y; \mathcal{G}) \arrow[rr, "d^{q-1}_{Y}"]
     &
      & C^q(Y; \mathcal{G}) \arrow[rr, "d^q_Y", shift left]
       &
        & C^{q+1}(Y; \mathcal{G}) \arrow[ll, "(d^q_Y)^{\ast}", shift left]
\end{tikzcd}
    \end{center}
    where $\Theta_{K,Y}^{q+1} = \{x \in C^{q+1}(Y,\mathcal{G}) : (\partial_Y^q)^\ast(x) \in C^q(K, \mathcal{F})$ and $\partial_{K,Y}^q$ is the adjoint of $\pi(\partial_Y^q)^\ast|_{\Theta_{K,Y}^{q+1}}:\Theta_{K,Y}^{q+1} \rightarrow C^q(K, \mathcal{F})$. We then define the q-th persistent sheaf Laplacian by:
    \begin{center}
        $\Delta_{K,Y}^q = (\partial^q_{K,Y})^\ast \partial^q_{K,Y} + \partial^q_K(\partial^q_K)^\ast$
    \end{center}
    The nullity of the q-th persistent sheaf Laplacian is then equal to the q-th persistent sheaf Betti number. The matrix representation of the persistent sheaf Laplacian is constructed similarly to the combinatorial Laplacian.

    \subsubsection{Protein-Protein Interaction Complexes}
    Given what the tensor factorization has identified as domain defining gene programme activity, we want to further investigate these programs and how they function at the protein level, and in particular what genes may be driving their function. To this end, we select the top weighted genes from a disease-associated tensor factor and consider its PPI network from the STRING database \cite{Szklarczyk2023}. A PPI network is an undirected graph where each gene is represented as a node and there exists a pairwise connection between two genes if they are either experimentally known to interact or exhibit significant co-expression. In terms of their biological significance, co-expressed genes are understood to be controlled by the same transcriptional regulatory program, or they may be functionally related or members of the same pathway or protein complex. Protein interacting genes code for proteins which interact, which is important for understanding pharmacological effects.

    Taking some confidence score of interaction as a prior given, we obtain the PPI network as an unweighted undirected graph $G = (V,E)$. We can then add weights to the edges based on the co-expression of those genes in a specific spatial domain or cell type of interest. The clique complex of $G$ is a simplicial complex whose simplices are all the cliques of $G$. A clique of $G$ is a subset of nodes where every single node is directly connected to every other node within that subset. In this sense, we are constructing, say, 2-simplices (triangles) from the sets of three mutually connected nodes, and so on. Considering computational cost as well as conceptual simplicity, we add simplices only up to 2 dimensions, so that we may consider the $L_0$ and $L_1$ Persistent Sheaf Laplacians. Filtration on such a complex then preserves those PPI interactions which are also shown to be strongly co-occurring in a specific disease setting. These interactions may be inferred as important drivers of some pathology.

    \subsubsection{Topological Perturbations}
Given our PPI clique complex, in which the genes make up the set of 0-simplexes, the co-expression weights facilitate the construction of an inverse rips filtration \cite{edelsbrunner2002persistence,su2025topological}. We wish to now identify the topologically significant genes via topological perturbations over each scale of the filtration \cite{cottrell2025repurposing, du2024multiscale}. To establish our sheaf, we must assign charges to each gene. These charges are taken from the factor analysis, so that the sheaf is scaled by the pathways whose usage moves with disease in the domain under study, i.e. emphasizing those genes which have been inferred as disease-associated in that domain. For a domain $d$, let $R_d$ be the set of programmes whose centred usage separates the conditions there, and let $\delta_r$ be the signed effect size of that shift. The analysis is carried out once for each $r \in R_d$: the factorization has decided which programmes define the domain and move with disease, and the topological question is which genes are structural within that programme.  For programme $r$, the charge of gene $g$ is $c^{(r)}_g = B_{gr}\,\delta_r$, the gene's loading on that programme weighted by how far the programme moves. Because $B$ is non-negative, the sign of this charge is inherited entirely from the programme, while its magnitude combines the strength of that participation with the size of the movement, so that within one network all charges share a sign and the loadings set their size. We retain the $300$ genes of largest loading on $r$ as the vertex set of that programme's network, so that the filtration is built over the part of the genome the disease contrast implicates. We hypothesize that the genes which are most topologically significant with respect to a pathway importance scaled sheaf would correspond well to significant biomarkers of disease progression. Topological significance is calculated according to the disparity between feature vectors obtained from the original complex and a perturbed complex where the respective gene has been removed.

Specifically, for a clique complex $K$, we obtain the sets of subcomplexes induced by filtration: $\{K_0, K_1, ..., K_p\}$. For each step of the filtration we compute the $L_0$ and $L_1$ Persistent Sheaf Laplacians. The spectra of these Laplacians encode the topological and geometric structure of our network. We remove a specific gene from the network to construct the perturbed complex $\hat{K_i}$. The spectra of the combinatorial Sheaf Laplacian at that scale give  us a feature vector for each subcomplex. Having obtained a feature vector for $K_i$- call it $f_i$, and $\hat{K}_i$- call it $\hat{f}_i$, we calculate the Wasserstein distance between the two: $\text{dist}(f_i,\hat{f}_i)$. The larger the distance, the more topologically significant we consider that gene to be at that scale. However, this can yield a notion of topological significance that is strongly linked to the degree centrality of that gene. While hub genes may well be biologically consequential, their prominence under this statistic largely restates their connectivity, which fails to capture certain genes of pharmacological significance such as those linking pathways. Furthermore, it biases the analysis toward already well studied genes. It is more interesting then to consider those genes whose removal disturbs the spectrum by more than their connectivity at that scale would predict. Thus, we regress $\text{dist}(f_i,\hat{f}_i)$ on the weighted degree of the removed gene in $K_i$, and rank genes by the residual rather than by the distance itself. The regression is carried out separately at every step of the filtration, and genes absent from a given scale are excluded from that fit rather than assigned a residual of zero. We then take as topologically significant those genes that remain at the top of the residual ranking at every scale.

\subsection{Agentic Panels}

The agentic panels proceed as a sequence of interpretations of the mathematical semantics product at each step. First, the agents attempt to infer from the tensor factorization, optimal transport plan, clustering results, and QC statistics which regions of the tissue are most relevant for a differential gene expression analysis of the diseased tissue. To this end, they have several motivations: to locate the tissue in which a disease effect is concentrated, to ensure that whatever region is chosen can actually support a statistical test, and to avoid regions whose apparent signal may be a property of the measurement or batch status rather than of the disease. For this reason, agents are encouraged to be more conservative. These motivations pull against one another, so the panel separates them. Five specialists each receive exactly one perspective of a candidate region, namely its spatial arrangement, its cellular composition, its programme usage, its transport correspondence with the other condition, and its design and quality characteristics, with the evidence packets filtered mechanically so that no specialist sees another's material. Each returns a relevance score and a suitability score, and must attach to every claim the packet field and value on which it rests. A critic then examines each cited item and labels it independent, redundant, explained away by a confound, or unverifiable, so that agreement between two specialists counts only when they are reading different things. A synthesis applies a stated rule to what survives: a region is shortlisted only if at least two independent lines of relevance from different perspectives remain, suitability passes, and no decisive confound is left unresolved; a single surviving line, or failing suitability, places the region on hold; anything else is excluded. Specialists may also propose comparisons other than the default case-versus-control contrast, such as a region against the tissue it maps onto in the other condition under the transport plan, a neighboring region within one condition, a pooling of regions, or an interaction between the two, and nominate the cell types in which each comparison is estimable; every proposal is judged by the critic before any test is run.

Based on a selection of disease-associated regions and cell types, the agents now must select which genes they infer to be worthwhile targets for pharmacological intervention. They approach this question in three ways, corresponding to the three independent lines of evidence the preceding analysis produces. One specialist reads the differential expression, asking whether a gene's transcript abundance separates the conditions in the selected regions and in which cell populations. A second reads the factorization, asking whether the gene carries weight in a programme whose usage moves with disease in those regions. A third reads the interactomic networks, one per programme upheld in the region, asking whether removing the gene disturbs the topology of that programme's module beyond what its connectivity alone would predict, and naming the programme whose network the claim is about. As before, each specialist sees only its own axis. A fourth agent, provided with literature search, then determines the intervention direction in three recorded steps: what the gene product does, which way the process it serves moves in disease, and therefore whether the target should be activated or inhibited. This separation is necessary because the direction of a useful intervention is not always the direction in which the gene's own expression moves; a negative regulator of a lost process should be inhibited to restore it, whichever way its abundance went. A critic audits every claim and every citation, and the final tier of each gene is recomputed mechanically from the data axes the critic upheld. A target for which no direction can be sourced is carried forward as direction-open.

\subsubsection{Blinding and Literature Verification}
Because the agents' conclusions are compared with those of the study that generated the data, the agents must not have access to that study to ensure a fair comparison. The only description of the data any agent receives is a data card of tissue, source technology, cohort table, prior list of cell types, and prior domain count. Every literature request an agent makes is scanned against a blocklist of the source publication's identifiers and reported findings before it is sent. Every literature statement behind a mechanism or a direction is then machine-checked against the searches actually made during the run and against PubMed to limit LLM hallucinations, so that a direction resting on a citation which does not resolve, or which does not support the statement attached to it, is not fixed, and the target is carried forward as direction-open.

    \section{Results}
\begin{figure}[H]
        \centering
        \includegraphics[width=\linewidth]{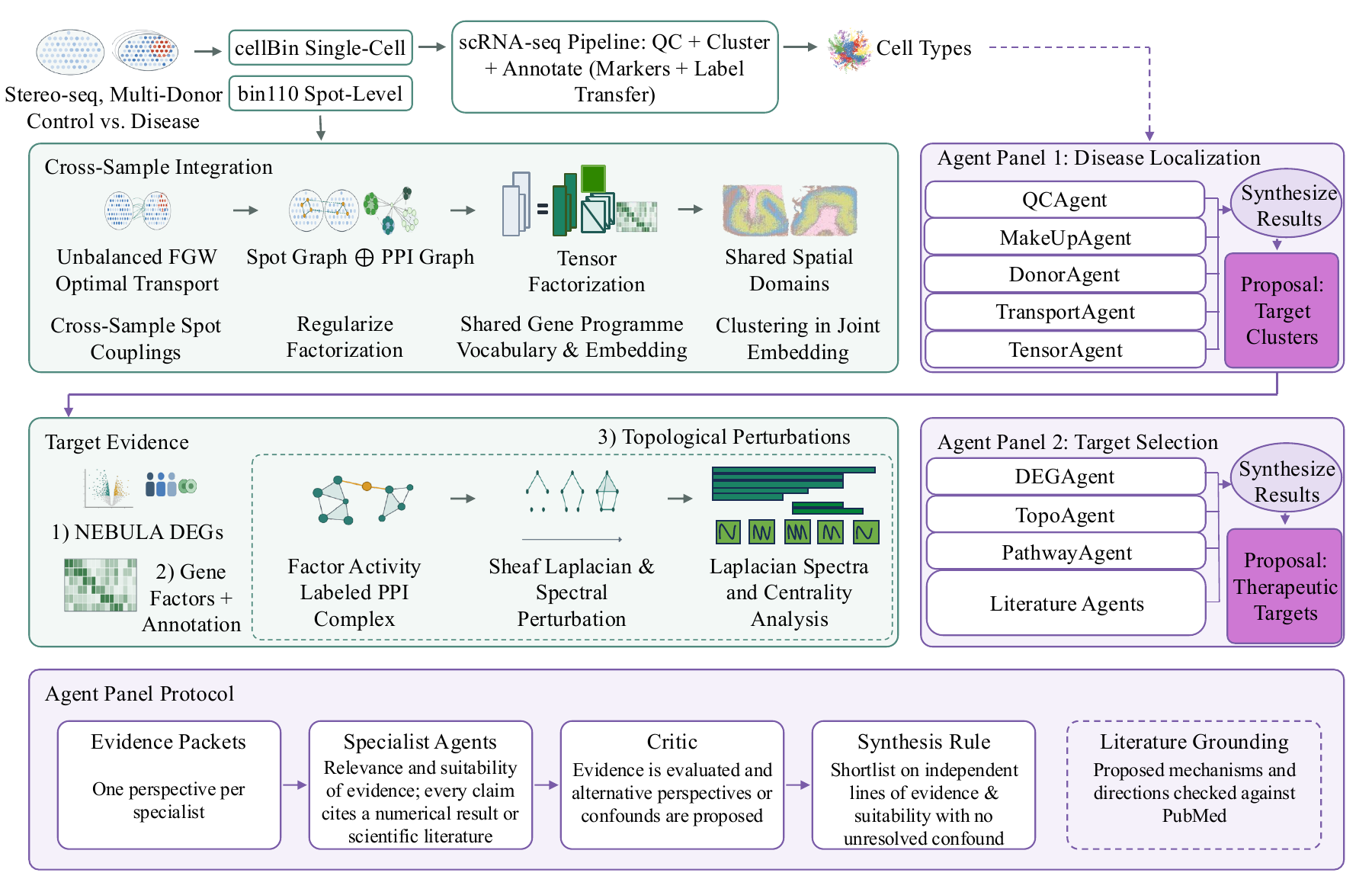}
        \caption{Overview of the TargetAgent pipeline. Multi-donor, multi-condition Stereo-seq sections are read at two resolutions: the cellBin view is carried through a single-cell pipeline to obtain cell types, and the bin110 view is integrated across samples for spatial domain analysis. Cross-sample integration couples spots between samples by unbalanced fused Gromov-Wasserstein optimal transport. It joins the transport couplings with a gene PPI graph to regularize the STORM irregular tensor factorization, and clusters the shared embedding into joint spatial domains with a common gene programme vocabulary. Agent panel 1 reads sequencing depth and batch confounds, cell-type composition, domain prevalence per donor, matched and unmatched transport mass, and programme factor activity to propose the disease-associated domains. Target evidence is then assembled within those domains from NEBULA differential expression, gene factor weightings with pathway annotation, and topological perturbation, via the persistent sheaf Laplacian, of one factor-charged PPI complex per disease-associated programme. Agent panel 2 reads the three axes in isolation, grounds mechanism and direction in the literature, and proposes therapeutic targets. In both panels, specialists cite only the model quantities in their own evidence packet, a critic audits every cited quantity, and a synthesis step applies a stated rule to what survives.}
        \label{fig:overview}
    \end{figure}

    TargetAgent is a proposed sequential mathematical pipeline to enable agentic modeling of spatially resolved omics data with the end goal of biomarker and therapeutic target predictions. In Figure~\ref{fig:overview} we give a broad overview of each of these steps. First, we seek to integrate data across disease conditions and donor cohorts. This is done by first constructing an optimal transport plan across samples to associate similar regions of tissue. Then, a graph regularized irregular tensor decomposition called STORM provides a joint embedding space for all samples so that spatial domains are directly comparable between disease statuses and patients. Simultaneously, the tensor decomposition provides a universal gene programme vocabulary which can be used to biologically interpret the activity within each cluster. Given these spatial domains, an agentic panel will consider which of them concentrates the disease localization or would be most relevant for downstream analysis of that disease, based on such criteria as measuring the un-transported mass, cell type compositions, gene programme activities, QC statistics, and more. Second, we seek to select those genes within the chosen regions whose behavior the models mark as disease associated, using three independent readouts: the scale of a gene's dysregulation between conditions, its weight in a gene programme whose usage moves with disease, and its structural position in the protein interaction network of that programme. A second agentic panel reads each readout in isolation, grounds every mechanistic statement in the published literature, and proposes which genes are worth pursuing and in which direction each should be moved. Throughout, the mathematical models are responsible for what is measurable and the agents are responsible for interpreting what it means; every claim an agent makes must name the model quantity or the verified publication it rests on.

We applied this methodology to the Stereo-seq survey of aging and Alzheimer's disease in human prefrontal cortex \cite{gong2025stereoseq}. The study profiled post-mortem BA10 tissue from twelve male donors, six Alzheimer's cases (ages 70--91, Braak IV--VI\cite{braak1991staging}, CERAD C2--3) and six age-matched controls (ages 74--95, Braak II--III, CERAD C0--1), with two cryosections per donor. The authors analyzed the array at bin110, at bin50, and at the cellBin level of individual segmented nuclei, of which 398,741 were retained after QC filtering. Cell types were called from the marker sets of Mathys et al.\ \cite{mathys2019} into seven classes: excitatory neurons, inhibitory neurons, astrocytes, oligodendrocytes, oligodendrocyte precursors, microglia, and endothelial cells. A spatially constrained clustering resolved seven tissue compartments: the six neocortical layers and the white matter, validated by layer markers including {AQP4} in L1, {HPCAL1} in L2/3, {PVALB} in L4, {PCP4} in L5, {KRT17} in L6, and {MBP} in white matter. The principal findings of the original study are the following. Layer architecture degrades with disease, with layers II--VI contracting and layer I expanding in severe AD alongside reactive astrocytes expressing {GFAP} and {NEAT1}. Downregulation of mRNA dominates upregulation in both the moderate and
severe AD cases, and the white matter carries the largest downregulation in moderate AD, involving synaptic transcripts ({SNAP25}, {S100B}, {CALM2}), and myelin maintenance ({MBP}) and {SLCO5A1}. {QDPR} peaks in moderate AD as a compensatory response, and {SPP1} and {SLC38A2} are highest in moderate and lowest in severe disease. Neuroprotective transcripts that healthy aging deploys near stressed neurons, such as {VGF}, {LMO4}, and {NRN1}, lose that pattern in AD, and instead a heat-shock and clearance response is initiated via ({HSP90AA1}, {HSPB1}, {CRYAB}, {SORT1}). Finally, the authors nominate the transcription factor {ZNF460} as a regulator of three neuronal co-expression modules that fall with disease progression.

In our work, none of this prior information was explicitly available to the pipeline: the agents are intentionally instructed not to search for or use the publication that produced the dataset so that their results would be un-influenced by the original study. To ensure this, every literature request is scanned against a blocklist of findings and identifiers before it is sent, and the only description of the data any agent receives is a data card of tissue, source technology, cohort table, prior list of cell types, and prior domain count. The comparisons we draw below between the
agents' conclusions and the authors' are therefore comparisons between two independent readings of the same tissue. One based on expert biological analysis, the other based purely on an LLM reading of certain mathematical semantics.

\begin{figure}[H]
        \centering
        \includegraphics[width=\linewidth]{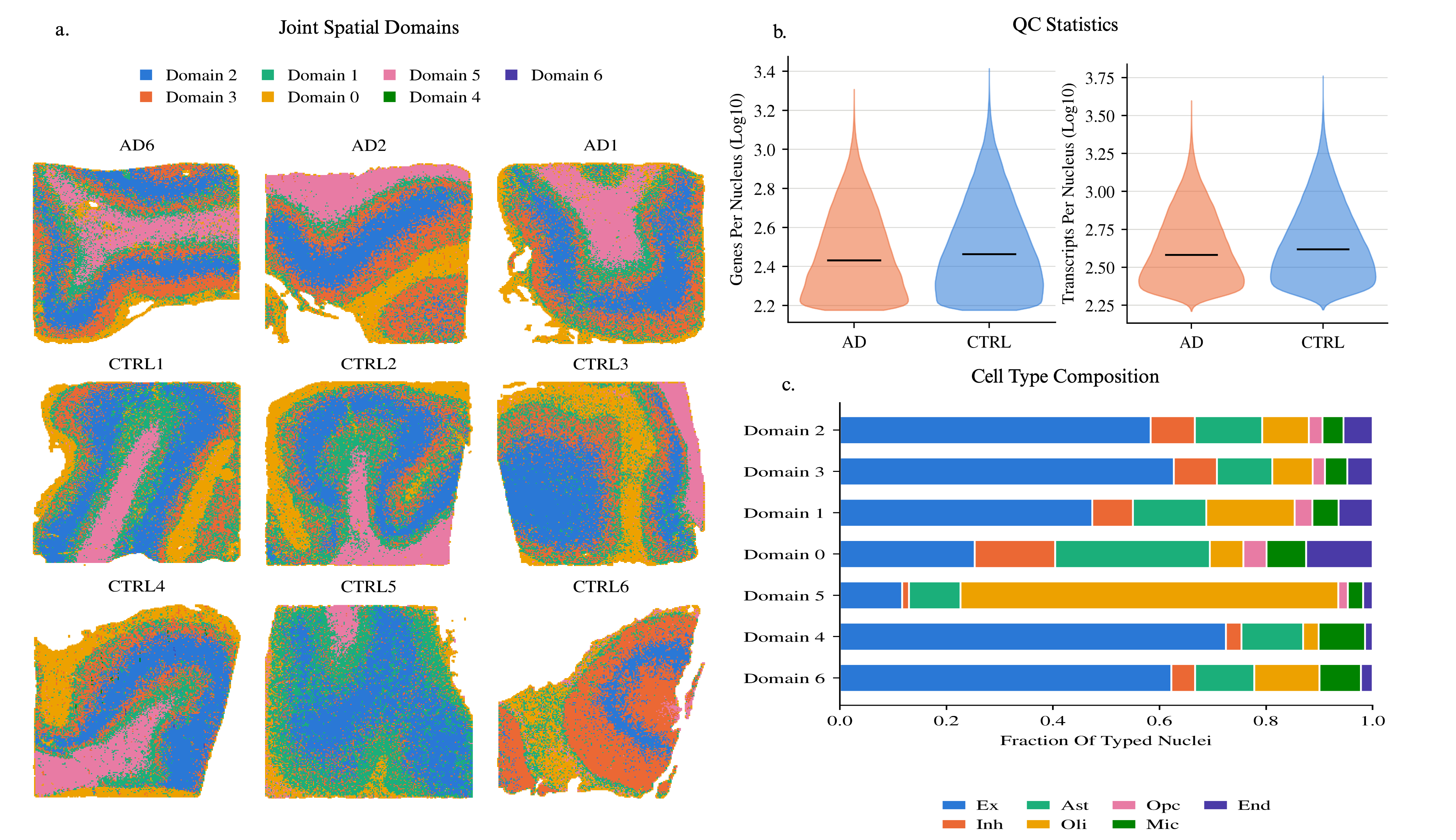}
        \caption{Joint spatial domains and cohort statistics for the nine retained sections. (a) The seven joint domains obtained by Leiden clustering of the shared STORM usage embedding, shown on the three moderate AD sections and the six control sections. Domain 5 is the white matter, domains 2 and 3 are the two grey matter compartments, domain 1 is the grey-white interface, and domain 0 is the astrocyte-rich marginal compartment; domains 4 and 6 hold under fifty tiles and are treated as nuisance. (b) Genes and transcripts per nucleus (log10) in the cellBin view by condition, with medians marked, showing the depth gap between AD and control that is carried as a covariate in every differential expression model. (c) Cell-type composition of each domain as the fraction of typed nuclei, from the consensus annotation of the cellBin view against the Mathys et al.\ vocabulary.}
        \label{fig:domains}
    \end{figure}

Our analysis uses only nine of these sections, one per donor: the three moderate-stage cases (Braak 4, 6 and 6) and all six controls. We excluded the three severe AD sections before fitting, on our own quality assessment rather than on any agent's advice, because their RNA content was depleted to the point of incomparability in a DEG analysis. Binning the retained sections into bin110 tiles with at least 50 counts gives 252,077 tiles. Segmentation yields 707,012 nuclei, of which 374,603 express at least 150 distinct genes and enter the single-nucleus pipeline: clustering the Harmony-corrected embedding gives 13 clusters, and the consensus of three independent callers resolves all of them with a cell type labeling, with all three callers agreeing on 9 clusters covering 86.2\% of nuclei and no nucleus left unassigned a biological label. The cell type vocabulary and its marker sets are, as in the source study, taken from Mathys et al., so that the two analyses are typed against a common reference. Composition is excitatory-dominated in every section, with excitatory fractions from 0.431 (AD2) to 0.567 (CTRL6) and oligodendrocyte fractions from 0.081 (CTRL6) to 0.263 (AD2) (Figure~\ref{fig:domains}c). The quality statistics themselves are group-dependent and this governs the differential expression reads (Figure~\ref{fig:domains}b). Median counts per tile are 3491, 5229, and 4190 in the three AD cases against 4828 to 7567 in the controls, and the fraction of segmented nuclei passing the 150-gene threshold is 0.356, 0.553, and 0.401 in AD against 0.522 to 0.740 in controls, giving a Spearman correlation between section depth and disease status of $-0.639$.
Batch structure is not orthogonal to group either: the GEF writer version S64\_i4 appears only in AD1 and AD2, and chip lot B01809 is the single lot shared across condition status. We note that the original authors read the same retention gap as biology, as evidence of cellular degradation and Wallerian-like degeneration; while this may be true, for the sake of DEG analysis for targeting we choose to treat it as a measurement confound to be carried explicitly, since in a three-against-six design the two readings cannot be separated. In practice this means per-section
depth enters every differential expression model as a covariate, programme movement is read from usage with each section centred on itself rather than from amplitude-carrying activity, and any compositional change is checked against the direction that differential retention would produce. This results in generally a more conservative reading from the agents, which was desirable to limit their attempts at biologically justifying what might be in part a technical confound.

Joint spatial domains were then obtained on the bin110 view by jointly clustering the shared STORM $Q_k H$ usage embedding across samples via Leiden \cite{traag2019leiden}, at the resolution whose domain count matches the assumption of six neocorticol layers and white matter as in Mathys et. al. The seven domains are shown for all nine sections in Figure~\ref{fig:domains}a and are banded rather than patchy, and all five substantial domains are present in every donor. Five are large enough to test: domain~2 (71,541 tiles, 28.4\% of the tissue, spatial coherence 0.738; Ex 0.58, Ast 0.13, Oli 0.09), domain~3 (59,422, 23.6\%, 0.555; Ex 0.63, Ast 0.10, Inh 0.08), domain~1 (57,014, 22.6\%, 0.478; Ex 0.47, Oli 0.17, Ast 0.14), domain~0 (35,135, 13.9\%, 0.709; Ast 0.29, Ex 0.25, Inh 0.15) and domain~5 (28,891,
11.5\%, 0.818; Oli 0.71, Ex 0.12, Ast 0.10); domains 4 and 6 hold 45 and 29 tiles. Domain~5 is unambiguous white matter and carries the highest spatial coherence in the partition, though it is fragmentary in the two thinnest control sections; domains 2 and 3 are the two grey-matter compartments, domain~1 is the grey--white interface, and domain~0 is an astrocyte-rich, low-depth compartment lying at the tissue margin. The STORM enabled partition therefore reproduces the grey--white axis and the marginal compartment of the original laminar segmentation. Compositional disease effects are confined almost entirely to the oligodendrocyte lineage, and they run upward: Oli rises 0.123 to 0.304 in domain~1 ($p=0.002$), 0.064 to 0.129 in domain~2 ($p=0.036$) and 0.043 to 0.113 in domain~0 ($p=0.036$), while excitatory neurons fall 0.605 to 0.541 in domain~2 ($p=0.006$).

The first agentic panel was then asked which of these domains is worth testing for differential expression between diseased and healthy donors. A panel of five specialist agents each received exactly one mathematical perspective: spatial arrangement, cell-type composition, programme usage, transport couplings, and design and quality statistics, a critic labeled every cited quantity independent, redundant, explained away, or unverifiable, and a synthesis applied a stated rule requiring at least two independent surviving lines from different perspectives, passing suitability and no unresolved confound. On the factor perspective the panel was shown all ten STORM gene factor loadings, their domain-level usage, and their case--control contrasts, and it cited four: the myelin programme F1 (PLP1, MBP, MOBP, CNP, MAG, CRYAB, BCAS1, ERMN), whose usage falls in domains 5, 3, and 0
($d=-2.13$, $-2.01$, $-2.05$); the neurofilament programme F8 (NEFM, INA, SCN1B, NEFH, NEFL), whose usage rises in domain~2 ($d=+2.29$); the synaptic programme F3, whose usage falls in domain~3 ($d=-2.21$); and the haemoglobin-dominated F4, whose fall in domain~0 the specialist agent flagged as tracking the depth confound. The critic upheld the F1 citations in domains 5 and 3, judged the F8 rise not independent of the oligodendrocyte influx because that programme's marker enrichment is itself oligodendroglial, and discarded F4 as a nuisance axis. On the transport perspective the panel read how much of each domain's tissue fails to find a counterpart in the other condition under the optimal transport plan. Domain~5 gives the clearest result and the panel's strongest single line: 0.300 of case-side mass is unmatched, and its self-mapping falls from 0.615 within condition to 0.381 across it, while the reverse direction is almost freely matched (unmatched 0.075, self-map 0.696). The inference is that AD-affected white matter therefore contains tissue that control white matter does not. This matches the findings of the original study regarding white matter carrying the most severe downregulation in moderate AD. Domain~0 shows a weaker version of the same asymmetry (0.233 unmatched case-side), domain~1 shows unmatched mass that is large but symmetric (0.194 against 0.187) and was therefore read by the agent as donor heterogeneity rather than disease, and domains 2 and 3 are largely conserved (0.109 and 0.127 case-side). Two domains were thus shortlisted for subsequent DEG analysis. Domain~5 ranked first on three surviving lines from three perspectives: its cluster size and structure across donors and conditions means it is able to carry statistical comparisons; the transport asymmetry above; and the F1 usage fall, which the panel argued cannot be produced by composition, since the oligodendrocyte fraction actually rises in disease, nor by amplitude in the STORM factorization, since F1's donor specific amplitude also rises due to the cell-type composition shift. Domain~3 ranked second on the F1 fall and an excitatory neuron fraction loss at $p=0.054$. The remaining three were held, and the reasons are as informative as the selections: domain~2 has the highest relevance in the panel and the best suitability but only one surviving line, its headline F8 statistic having been judged non-independent by the agents due to potential lateral RNA diffusion; domain~1 carries the largest compositional effect in the cohort but fails suitability on group-dependent fragmentation; and domain~0's endothelial deficit was explained away by the agent as selection against RNA-poor nuclei. We again note that the panel arrived at white matter as the seat of the moderate-disease signal from geometry, transport, and programme usage alone, which is where the original authors also locate the largest moderate-AD transcriptional change, although they reach it from a layer-stratified differential expression analysis and expert annotation that the panel never saw.

    \begin{figure}[H]
        \centering
        \includegraphics[width=0.9\linewidth]{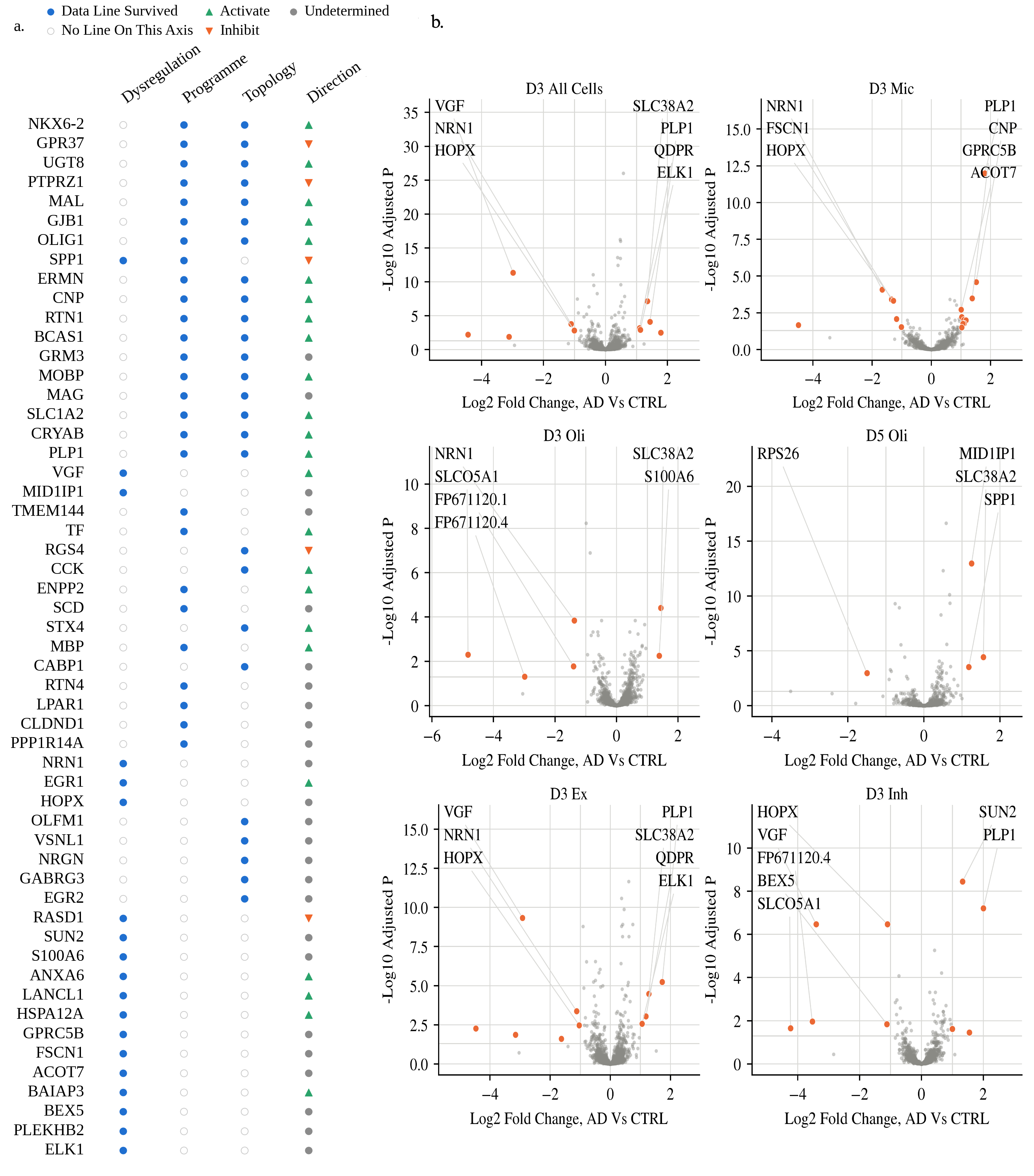}
        \caption{Target ranking and differential expression in the shortlisted domains. (a) The admitted targets from the second agentic panel, ordered by tier, with the three data axes on which a line of evidence survived the critic (dysregulation, programme membership, and topological significance) and the intervention direction fixed by the literature agent, where a grey marker denotes a target carried forward as direction-open. (b) NEBULA volcano plots for the whole-domain and cell type-stratified contrasts of AD against control in domains 3 and 5, with hits at FDR $<0.05$ and $|\log_2\mathrm{FC}|>1$ labeled. Per-section depth enters each model as a covariate and donor section as a random intercept.}
        \label{fig:targets}
    \end{figure}

Within these two domains the second panel produced a ranked target list, summarized with its supporting axes in Figure~\ref{fig:targets}a. The candidate pool is the union over the three axes of every gene called by differential expression in a contrast with a shortlisted domain on its side of interest, every gene in the topological significance set of any programme network, and the twenty genes of largest weighting in a disease-associated factor, which gives 101 gene and domain pairs over 73 distinct genes. The panel placed 16 in a convergent tier, 13 as single-axis with a strong mechanism, 25 on hold, and 19 rejected, so that 29 targets were proposed and 54 in total were admitted for consideration; the 25 on hold were carried as direction-open because no verified literature source fixed a desired direction for treating AD, and the critic raised a disqualifying concern against 17 candidate entries. A module leads the set: NKX6-2, GPR37, UGT8, PTPRZ1, MAL, GJB1, OLIG1, ERMN, and BCAS1 are each regulators or effectors of oligodendrocyte differentiation and myelin assembly, and every one of them carries the programme and topology lines in both domains except PTPRZ1 and BCAS1, whose topology line holds in the white matter only. Within the convergent tier also sit SPP1, the one target for which the two surviving lines come from different measurement units, a donor-level differential expression call and a factor weighting, and which is assigned {inhibit} on the reasoning that it restrains OPC progression; the negative regulators GPR37 and PTPRZ1, assigned {inhibit} on the reasoning that restoring a lost process means blocking what inhibits
it; and the structural myelin genes CNP, MOBP, and PLP1, which are admitted but explicitly discounted by the agents as they do not drive pathology, with MBP admitted on the programme axis alone and MAG held because its periaxonal and neurite-inhibitory roles argue opposite directions. Considering PPI topology as a statistic orthogonal to loading rankings places NKX6-2, GPR37, UGT8, MAL, and GJB1 above the structural genes. The neuronal targets are carried on single axes: VGF on its differential expression, and RGS4, CCK, and STX4 on their positions in the neuronal programme PPI network of domain~3; NRN1 is held because the valence claim behind its direction was filed without a citation. Twenty of the admitted targets were flagged by the agents as novel in this indication in the context of AD, seven among the proposed targets, NKX6-2, UGT8, GPR37, PTPRZ1, GJB1, ERMN, and BCAS1, and thirteen on hold, among them TMEM144, CLDND1, PPP1R14A, GRM3, and MID1IP1. Every literature statement behind a direction was machine-checked against the searches made during the run and against PubMed to limit LLM hallucinations \cite{ji2023hallucination}.

The first axis is differential expression, which we computed with Nebula over the nine domain / cell-type contrasts recommended by the previous panel \cite{he2021nebula}. At FDR $<0.05$ and $|\log_2\mathrm{FC}|>1$ this returns 65 hits over 29 distinct genes (Figure~\ref{fig:targets}b). The largest and best-estimated effect is the loss of VGF in cortex ($-2.98$ whole-domain, $p_{\mathrm{adj}}=4.8\times10^{-12}$; $-2.92$ in excitatory and $-3.41$ in inhibitory neurons), accompanied by NRN1 ($-1.11$) and HOPX ($-1.00$); the largest white-matter effect is a rise in MID1IP1 ($+1.22$, $p_{\mathrm{adj}}=7.4\times10^{-21}$); and SPP1 and QDPR rise in both compartments (SPP1 $+1.79$ in cortex, $+1.21$ and $+1.18$ in white matter; QDPR $+1.10$ to $+1.25$ across four cortical contrasts). Four of these agree with the original study in both identity and direction. The authors report QDPR peaking in moderate AD as a compensatory response and SPP1 highest in moderate disease; we recover both, in the same stage of disease, though the panel declined QDPR as a target, on the grounds that inhibiting dihydropteridine reductase reproduces BH4 deficiency and that a rise of similar size in three unrelated lineages also fits a section-level difference, which is a judgement about intervention rather than about the measurement. They report that the neuroprotective programme of healthy aging, VGF and NRN1, is lost in AD; we recover the loss of both as our two largest neuronal effects. Two hits diverge sharply in interpretation rather than in measurement. SLC38A2 rises and SLCO5A1 falls in essentially every stratum of both domains, and the original authors report both as disease biology, SLC38A2 as a cellular stress response peaking in moderate AD and SLCO5A1 among the transcripts most downregulated in moderate-AD white matter. The agentic panel's critic instead rejected both: SLC38A2 as a canonical amino-acid-deprivation gene rising uniformly in every cell type in a three-against-six design with no post-mortem interval or RNA integrity data, and SLCO5A1 for a near-constant $-4.4$ log fold change across every stratum with borderline adjusted p-values, which it read as a per-section detection artifact. We do not claim the agents are right and the authors wrong; we note that a uniform, compartment-independent shift is exactly the signature the panel was instructed to distrust. The panel also read the myelin transcripts against itself: PLP1 and CNP rise in astrocytes, excitatory neurons, inhibitory neurons, and microglia but not in the oligodendrocytes that make them. The panel attributes this to lateral spillover of abundant myelin transcripts into neighboring segmented nuclei, which is common in Stereo-seq measurements, rather than to biology. The consequence is that no myelin structural gene shows a measured case--control decrease anywhere in this cohort, whereas the original authors report MBP downregulation in moderate-AD white matter. We attribute this divergence primarily to the statistical test rather than to the tissue. The original analysis applied a two-sided t-test to individual pseudo-spots and nuclei, with a fold-change threshold of 0.25 \cite{gong2025stereoseq, wolf2018scanpy}. That test treats every spot as an independent observation, so differences between donors enter as if they were replicated thousands of times, and it carries no term for sequencing depth, which in this cohort is strongly correlated with disease status. Our model is deliberately stricter on both counts: the donor section enters as a random intercept, so that the effective sample size is the number of donors; per-section depth enters as a covariate, so that a depth difference between groups is not immediately read as downregulation; and a hit must exceed $|\log_2\mathrm{FC}|>1$. Under these conditions an apparent loss of abundant transcripts such as MBP in lower-depth AD sections is absorbed by the donor and depth terms. We therefore read the four concordant genes, VGF, NRN1, SPP1, and QDPR, as the effects robust to the choice of model, and the myelin structural genes as effects that do not survive adjustment for donor and depth. This means the myelin story in the agentic analysis rests on the tensor factor programme and PPI network axes rather than on differential expression.

    \begin{figure}[H]
        \centering
        \includegraphics[width=\linewidth]{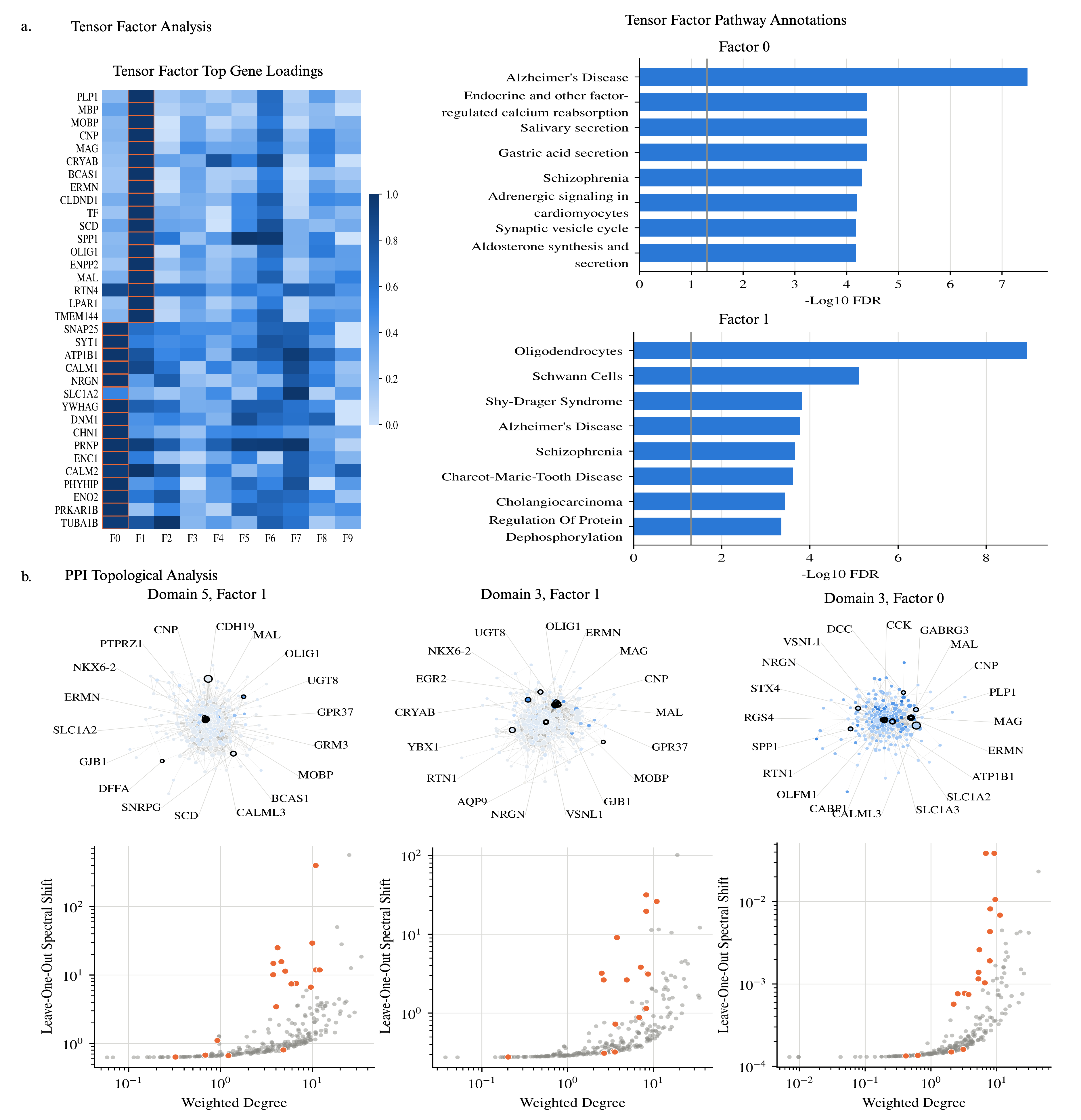}
        \caption{Programme and topological evidence for the shortlisted domains. (a) Top gene loadings of the rank-10 STORM gene factor matrix $B$, with the highest-loading genes of the myelin programme F1 and the synaptic programme F0 outlined, alongside the over-representation of each programme's fifty highest-loading genes against the pathway, cell-type marker, and disease gene sets described in Methods, with the FDR $=0.05$ line marked. (b) PPI topological analysis, with one network per programme upheld in a domain: the F1 network of domain 5, and the F1 and F0 networks of domain 3. Also shown is the clique complex over that programme's 300 highest-loading genes, with vertices colored by the signed charge and the topologically significant genes labeled. The leave-one-gene-out Wasserstein shift of the persistent sheaf Laplacian spectrum is regressed against weighted degree, with the genes retained at the top of the degree-corrected residual ranking at every filtration scale highlighted.}
        \label{fig:storm}
    \end{figure}

The second axis comes via the STORM factorization (Figure~\ref{fig:storm}a). The rank-10 programme vocabulary recovers the cellular architecture of the prefrontal cortex: marker scoring assigns factors F0 (SNAP25, SYT1, ATP1B1, CALM1, NRGN) and F2 (HSPA8, NDUFA4, PPIA, UQCRH, COX6C) to excitatory neurons, F1 and F6 to oligodendrocytes, F7 (SLC1A2, SLC6A1, SLC1A3, PTN, PTPRZ1) to astrocytes, and F4 (HBA2, HBB, HBA1, CLDN5, IGFBP7) to the vascular and erythroid compartment. Pathway annotation via over-representation of each programme's fifty highest-loading genes reproduces that reading independently and connects it to disease: F1 is enriched for Oligodendrocytes and Schwann Cells and for Alzheimer's Disease, while F0 is enriched for Alzheimer's Disease, the synaptic vesicle cycle, and negative regulation of amyloid-beta formation. The spot-level loadings place F0 in the two cortical compartments and F1 in the white matter, which is the same grey--white separation the original authors obtain from layer markers. Across programmes and domains with only nine samples no shift reaches an FDR of 0.05, and the smallest, 0.150, is attained by the F1 usage fall in domains 5, 3, and 0, by the F8 rise in domain~2, and by the F4 fall in domain~0. In spite of lacking statistical significance, the agent proposes three properties of the F1 result as to why it nonetheless should carry an analysis. It is the largest absolute within-section shift in the fit, a 28\% reduction in the programme that defines the white matter. It is counter-directional to both alternatives that would otherwise explain it, since the oligodendrocyte composition actually rises in AD rather than falling, and since F1's cohort amplitude in the $D_k$ matrix also rises while its usage falls. And it recurs in three separate compartments. To this end, the agents still propose some significance to this observation. The reading the panel takes from this is that oligodendrocytes in moderate AD are more numerous but individually less engaged in the myelin programme, which is a state change rather than a cell loss. This is where the agentic analysis and the original authors' converge most strongly at the level of mechanism and diverge at the level of evidence: they report white matter as the compartment with the largest moderate-AD downregulation, including myelin maintenance genes, from per-gene tests, whereas our per-gene DEG tests using Nebula find no myelin decrease at all due to stricter modeling constraints. The signal then appears only once expression is decomposed into programmes and read as usage across the tissue.

The third axis then asks which genes are structurally central to a gene programme rather than merely strongly loaded in it in the $B$ matrix (Figure~\ref{fig:storm}b). The factorization decides which programmes define a domain and move with disease there, and the topological analysis then asks which genes are structural within that programme. The network vertices are that programme's 300 highest-loading genes, charged with their loading times the programme's shift, with STRING curated PPI interactions supplying the edges and co-expression over the domain's cells their weights, and a gene loading on two upheld programmes is a vertex of both networks and is assessed separately in each. On each network we compute the leave-one-gene-out shift of the persistent sheaf Laplacian spectrum to measure a genes topological significance in the programme wide PPI structure. The statistic we take forward is the residual after weighted degree is regressed out at each filtration scale separately, and a gene qualifies only if it stays at the top of that residual ranking at every scale in which it appears. Thus, a gene is important for structural significance that goes beyond hub centrality, to help de-emphasize the bias toward well studied genes and emphasize novel markers. Domain~5 carries one upheld programme, F1, whose network has 275 of its 300 genes in the interaction graph and 1,534 edges, and it returns 18 genes: MOBP, CNP, BCAS1, ERMN, OLIG1, MAL, NKX6-2, GPR37, GRM3, SLC1A2, UGT8, GJB1, PTPRZ1, CDH19, SNRPG, SCD, DFFA, CALML3, of which 13 hold the top of the residual across all filtration scales. Domain~3 carries two upheld programmes, and so two PPI networks. Its F1 network shares the vertices and edges of the white matter network but is weighted by the cortical domain's own cells, and returns 17 genes: CNP, MAG, ERMN, MAL, VSNL1, GPR37, NRGN, RTN1, UGT8, NKX6-2, MOBP, OLIG1, GJB1, CRYAB, YBX1, EGR2, AQP9. Its F0 network, over the neuronal programme, has 256 genes in the graph and 1,130 edges and returns 20: SLC1A2, SLC1A3, VSNL1, NRGN, PLP1, RTN1, CNP, MAG, ERMN, ATP1B1, CCK, MAL, OLFM1, RGS4, CABP1, CALML3, STX4, GABRG3, SPP1, DCC. All three are known modules rather than gene lists. Seven genes hold the residual in both domain~3 networks, CNP, MAG, ERMN, MAL, VSNL1, NRGN, and RTN1, which is the strongest form of the claim, and the separation is informative in the other direction as well: MOBP and OLIG1 qualify in F1 and fail in F0, so their residuals are programme-specific rather than a generic property of the myelin block. However, PLP1 reverses this pattern: it is excluded from F1, where it is the principal hub at spectral rank 1, and included in F0 with high betweenness, at the interface between neuronal and myelin transcripts. The genes most strongly promoted by this perspective are the transcription factors NKX6-2 and OLIG1, the myelination-regulating receptor GPR37, the galactosylceramide synthase UGT8, the gap junction protein GJB1, and BCAS1, which marks actively myelinating oligodendrocytes. 

Taken together, the axes converge on a single module. The differential-expression axis captures the neuronal losses of VGF and NRN1 and the increases in SPP1 and MID1IP1; the factor axis captures the decline in myelin-programme usage; and the topological axis identifies the regulators and effectors of oligodendrocyte differentiation—NKX6-2, OLIG1, GPR37, UGT8, GJB1, and BCAS1—which are structurally central to the programme beyond their connectivity. This is the module on which the second panel's convergent tier rests, reached through three mathematical views of the same tissue, each interpreted by a specialist who had access to none of the others.

    \section{Discussion}

In this study, we have integrated a sequential mathematical analysis of omics data based on complementary mathematical perspectives such as optimal transport, tensor algebra, and topology with agentic AI to automate a co-scientist pipeline which takes us from multi-donor multi-condition spatial transcriptomics data to molecular target identification. We demonstrate this
procedure by applying it to a public Stereo-seq study of the Alzheimer's disease-affected prefrontal cortex and demonstrate that the agentic interpretations of the mathematical output are able to autonomously reach a consensus with several of the primary findings from biological expert analysis. While the original authors grounded their result in differential expression analyses on each tissue, in our case the results arose naturally from the structure of a tensor factorization which LLM agents were able to interpret. The results also show why the perspectives are complementary rather than redundant. The transport plan alone located the white matter as the compartment where diseased tissue has no healthy counterpart, with 0.300 of its case-side mass unmatched; the factorization found the fall in myelin programme usage and separated it from the rising oligodendrocyte composition that would otherwise explain it; and the topological analysis of the PPI network separated the regulators of the programme, NKX6-2, OLIG1, GPR37, and UGT8, from the structural genes that define it but do not drive it, which neither a loading nor a differential expression statistic can do in isolation. The agents' role was to read each of these where it is informative and to decline it where it is not, and the critic's rulings, with 45 cited quantities explained away and 21 unverifiable, are the record of that. The topological analysis of the gene interactions which comprise the tensor factors ultimately enabled the agents to propose twenty-nine therapeutic target hypotheses for addressing declining myelin production in AD-affected white matter, and flagged twenty genes as novel in this indication. Where the agents and the original authors diverge, as for SLC38A2 and SLCO5A1, the divergence is recorded together with the reasoning behind it, so that a reader can weigh the two readings against one another rather than take either on trust.

The proposed targets also hold up against literature the pipeline was not built on. The premise that oligodendrocytes and myelin are a therapeutic axis in AD rather than a bystander is now supported from several independent directions: single-nucleus profiling of the AD prefrontal cortex reported myelination genes among the most recurrently perturbed across cell types \cite{mathys2019}, a disease-associated oligodendrocyte state shared across AD models and other CNS pathologies has since been described \cite{kenigsbuch2022}, enhancing myelin renewal pharmacologically restores cognition in an amyloid mouse model \cite{chen2021}, and myelin dysfunction on its own is sufficient to drive amyloid deposition \cite{depp2023}. The genes at the head of the list are the recognized control points of that process rather than its bulk constituents: NKX6-2 is required for the normal structure of CNS myelin paranodes \cite{southwood2004}, OLIG1 is required to repair demyelinated lesions \cite{arnett2004}, and GPR37 is a negative regulator of late oligodendrocyte differentiation whose deletion accelerates myelination \cite{yang2016}, so the panel's activate and inhibit assignments carry the sign each gene has in that literature. PTPRZ1 is the case where the literature is split, since its phosphatase activity restrains differentiation while loss of the receptor impairs functional recovery from demyelination \cite{harroch2002}, and the panel's inhibit call should be read with that qualification. Beyond the myelin module, the proposed SPP1 target has recently been placed upstream of microglial synapse engulfment in AD mouse models \cite{deschepper2023}, and VGF was identified as a key causal regulator of AD by multiscale network modeling of an independent human cohort \cite{beckmann2020}. The agents were blocked from the source study, and while some of these references were reached by the direction agent's own audited searches, the rest were not. Either way, the point is that the targets the mathematics promoted from the structure of a single cohort are the control points the field has identified by other means over several years, which is what we take as evidence that the co-scientist is doing real work and that the mathematics is revealing something meaningful about the structure of the disease in this donor cohort. In this respect, we conclude that co-scientist frameworks have the capacity to drive scientific discovery when integrated with mathematical semantics to study omics data. 

We note two limitations of the current study. The first and most important is statistical. The analysis rests on nine sections from three moderate-AD and six control donors, and at this cohort size no change in programme usage reaches significance: the smallest false discovery rate, attained by the fall in myelin programme usage, is 0.150, and per-gene differential expression detects no decrease in any myelin structural gene. The central reading of this study, that oligodendrocytes in moderate AD are more numerous but less engaged in the myelin programme, is therefore a hypothesis consistent with the data rather than a statistically established finding, and the same holds for the proposed targets, their intervention directions, and the genes flagged as novel. Their agreement with the published literature shows that they are plausible, and thus convincingly presents this work as a proof of concept: it shows that agents reasoning over mathematical readouts can carry a spatial transcriptomics cohort from raw data to ranked, auditable target hypotheses. Recommendations reliable enough to act on will require larger donor cohorts, in which these effects can be tested with adequate power and replicated, along with additional perspectives measured independently of the transcriptome. The second limitation is the focus solely on the Stereo-seq platform. Imaging-based technologies rely on targeted gene panels, which biases target selection \cite{chen2015merfish}, so we restricted attention to whole-transcriptome technologies, and the subcellular resolution of Stereo-seq allows cell types to be annotated from the nuclear stain and carried into the differential expression analysis, which addresses the confound of cell-type composition. Lower resolution data such as 10X Visium could serve the same end through deconvolution against a single-cell reference \cite{im2023deconvolution}. Untargeted subcellular technologies do, however, suffer from weak capture efficiency and dropout, so a meaningful target may be expressed but missed, which denoising or imputation techniques may partly address \cite{cheng2023imputation}.

The perspectives considered in each agentic panel are is also a natural place to further explore and expand in future work. Each specialist is a perspective on one object. Nothing prevents adding more, or re-arranging them, to improve the scientific accuracy and motivation for each recommendation. A morphology agent reading the paired H\&E or ssDNA image would supply the histological context the transcriptome cannot, letting the panel weigh plaque and tangle burden, laminar boundaries, and tissue quality directly rather than inferring them from counts. A cell--cell communication perspective, built on spatially constrained ligand--receptor inference over the same domains, would test exactly the kind of finding our current panel structure cannot reach. One of the original authors' strongest mechanistic claims, the progressive loss of inter-layer signaling in moderate disease, has no counterpart among our five perspectives, and adding it would give the panel an independent line of evidence. Finally, the present study ends at a ranked list of hypothetical targets and the direction in which each should be moved. The translation of those targets into therapeutic modalities and candidate compounds is a task of its own, with its own sources of evidence and its own agentic reasoning, and is the subject of forthcoming work. Together these would greatly extend the principle the present study is built on.

\section{Acknowledgement}
This work was supported in part by NIH grant R35GM148196,   the University of Georgia, and the Georgia Research Alliance.

\section{Data and Code Availability}
The Stereo-seq AD-affected prefrontal cortex data can be obtained from the Gene Expression Omnibus under accession ID \href{https://www.ncbi.nlm.nih.gov/geo/query/acc.cgi?acc=GPL2948}{GSE269906}. The TargetAgent code is available in the GitHub repository at \href{https://github.com/seanfcottrell/TargetAgent}{https://github.com/seanfcottrell/TargetAgent}.

\section{AI Usage Declaration}
Throughout this work,  AI was used to assist with English editing. All written work was initiated, overseen, and approved by the authors. Claude (Anthropic) was used in part to help write and debug Python code for data analysis. The authors verified all outputs and results manually against the raw data. The AI agents referenced in this work were run using Claude Opus-5. 

\newpage
%\bibliographystyle{unsrt}
% \bibliographystyle{abbrv}
%\myexternaldocument{supplementary}

 \bibliographystyle{plain}
\bibliography{refs}

\end{document}